\documentclass[11pt,a4paper]{article}

\usepackage{jheppub}

\usepackage{amsmath}
\usepackage{amsthm}
\usepackage{amscd}
\usepackage{amsxtra}
\usepackage{upref}
\usepackage{amsfonts}
\usepackage{amssymb}
\usepackage{bbold}
\usepackage{graphicx}

\usepackage{pifont}

\numberwithin{equation}{section}

\title{Timelike U-dualities in Generalised Geometry}
\author{Emanuel Malek}
\affiliation{Department of Applied Mathematics and Theoretical Physics, \\Centre for Mathematical Sciences, University of Cambridge, \\ Wilberforce Road, Cambridge CB3 0WA, United Kingdom}
\emailAdd{E.Malek@damtp.cam.ac.uk}

\abstract{We study timelike U-dualities acting in three and four directions of 11-dimensional supergravity, which form the groups $SL(2)\times SL(3)$ and $SL(5)$. Using generalised geometry, we find that timelike U-dualities, despite previous conjectures, do not change the signature of the spacetime. Furthermore, we prove that the spacetime signature must be $\left(-,+,\ldots,+\right)$ when the U-duality modular group is either $\frac{SL(2)\times SL(3)}{SO(1,1)\times SO(2,1)}$ or $\frac{SL(5)}{SO(3,2)}$. We find that for some dual solutions it is necessary to include a trivector field which is related to the existence of non-geometric fluxes in lower dimensions. In the second part of the paper, we explicitly study the action of the dualities on supergravity solutions corresponding to M2-branes. For a finite range of the transformation, the action of $SL(2)\times SL(3)$ on the worldvolume of uncharged M2-branes charges them while it changes the charge of extreme M2-branes. It thus acts as a Harrison transformation. At the limits of the range, we obtain the ``subtracted geometries'' which correspond to an infinite Harrison boost. Outside this range the trivector field becomes non-zero and we obtain a dual solution that cannot be uniquely written in terms of a metric, 3-form and trivector. Instead it corresponds to a family of solutions linked by a local $SO(1,1)$ rotation. The $SL(5)$ duality is used to act on a smeared extreme M2-brane giving a brane-like solution carrying momentum in the transverse direction that the brane was delocalised along.}


\begin{document}

\hfill  DAMTP-2013-71

\vskip 1.5cm 
\maketitle
\flushbottom

\newcommand{\lb}{\ensuremath{\mathopen{<}}}
\newcommand{\rb}{\ensuremath{\mathclose{>}}}
\newcommand{\ph}{\ensuremath{\phantom}}
\newcommand{\ov}{\ensuremath{\overline}}
\newcommand{\og}{\ensuremath{\bar{g}}}
\newcommand{\hg}{\ensuremath{\hat{g}}}
\newcommand{\cg}{\ensuremath{\check{g}}}
\newcommand{\gm}{\ensuremath{\mathcal{H}}}

\newcommand{\be}{\begin{equation}}
\newcommand{\ee}{\end{equation}}

\newtheorem{defn}{Definition}[section]
\newtheorem{prop}[defn]{Proposition}
\newtheorem{thm}[defn]{Theorem}
\newtheorem{cor}[defn]{Corollary}
\newtheorem{claim}[defn]{Claim}
\newtheorem{rmk}[defn]{Remark}
\newtheorem{notation}[defn]{Notation}

\theoremstyle{definition}
\newtheorem{ex}[defn]{Example}

\section{Introduction} \label{SIntro}
String and M-theory, in addition to the usual 10- and 11-dimensional
Poincar\'{e} symmetry, contain a group of non-perturbative ``stringy''
symmetries, the so-called T- and U-dualities. These arise when studying
compactified backgrounds because the extended nature of the fundamental objects
-- strings and branes -- allows them to wrap the compact dimensions. In 10- and
11-dimensional supergravity, the low-energy descriptions of string and M-theory,
these symmetries manifest themselves through a group of global non-compact
symmetries of the lower-dimensional theories obtained by dimensionally reducing
along Killing vectors \cite{Cremmer:1977tt,Cremmer:1978ds,Cremmer:1979up}. These
symmetries generate transformations, linking different ``dual'' solutions which
from the perspective of string / M-theory are equivalent.

We study the action of U-dualities by using generalised geometry \cite{Gualtieri:2003dx,Hitchin:2004ut,Hitchin:2005in,Hitchin:2005cv,Hull:2007zu,Berman:2010is,Berman:2011pe,Berman:2011jh,Berman:2011cg,Berman:2012vc,Berman:2012uy,Aldazabal:2010ef,Pacheco:2008ps,Coimbra:2011ky,Hull:2006tp,Hull:2009mi,Hull:2009zb,Hohm:2010pp,Hohm:2010jy,Hohm:2012gk} which makes the duality symmetries of the supergravity manifest. The dualities then do not arise from dimensional reduction but rather form the inherent symmetries of the theory. We ultimately wish to make the $E_{11}$ symmetry manifest, as it has been conjectured that it is the underlying symmetry group of 11-dimensional supergravity \cite{deWit:1986mz,West:2001as,Riccioni:2007ni,Kleinschmidt:2003jf,West:2003fc,West:2004kb,West:2004iz,Riccioni:2006az,Cook:2008bi,West:2010ev,West:2010rv,West:2011mm,West:2012qz} and so as a first step we begin by restricting the dualities to act only in three and four ``dualisable'' dimensions. In \cite{Malek:2012pw} we discussed the action of U-dualities along three and four spacelike directions of 11-dimensional supergravity. Because we do not require a compactification in order to give rise to the duality symmetry, one may wish to dualise along time as well. After all, in order to construct the 11-dimensional supergravity as a non-linear realisation of $E_{11}$ we must allow dualities along time. In this paper we study this question by including time amongst three and four dualisable directions, thus paving the way for the construction of a non-linear realisation of $E_{11}$.

Previous works \cite{Hull:1998vg,Hull:1998ym,Hull:1998br} have studied the action of T- and U-dualities along timelike directions. There it was found that type IIA and type IIB string theories are related not to each other by T-dualities but rather to two different string theories, the so-called type IIA$^*$ and type IIB$^*$ theories, both of Lorentzian signature but with the ``wrong'' signs for the kinetic terms of the R-NS and R-R forms. The change of sign for the kinetic terms has been observed in Double Field Theory as well \cite{Hohm:2011zr,Hohm:2011dv}. Similarly, M-theory and its low-energy effective action, 11-dimensional supergravity, would not be invariant under the timelike dualities but would change signature, becoming the M$^*$ and M$'$ theories, containing various timelike directions. In \cite{Malek:2012pw}, we studied examples of dualities where we had Wick-rotated M2-branes to obtain a Euclidean worldvolume along which we can dualise. Wick-rotating back afterwards, we found Lorentzian solutions which exhibited some difficulties, for example complex or even singular metrics. We suggested in \cite{Malek:2012pw} that these difficulties arise because we are implicitly dualising along time and that when done explicitly, without Wick-rotation, these problems should disappear.

Here we will reinvestigate timelike dualities and see how generalised geometry deals with the problems that arise. We will briefly review generalised geometry in section \ref{SGenGeom}, before explaining how the metric and 3-form arise in the Euclidean generalised metric. We will show that these arise from a specific parameterisation of the generalised vielbein and that in the Euclidean case any vielbein can be brought into this form so that the description in terms of a generalised metric is always equivalent to the conventional one using a spacetime metric and 3-form. However, after we construct the ``Lorentzian'' generalised metric for when time is a dualisable directions in section \ref{STGenGeom}, we will show that this generalised metric is in general no longer equivalent to a spacetime metric and 3-form. Instead, the generalised metric can be of four types. Two of these can always be expressed in terms of a spacetime metric and 3-form but the other two need the inclusion of another bosonic field, the trivector $\Omega_3$ with components $\Omega^{ijk}$ which are totally antisymmetric. We will then, in section \ref{STU}, review the argument of how multiple timelike directions seem to appear when dualising along time before showing that this does not occur when we use generalised geometry. We also prove that the spacetime metric arising in the generalised metric will always have signature $\left(-, +, \ldots +\right)$. In section \ref{STDuality} we find the transformation laws for the bosonic fields for the three-dimensional case and we will see that a solution depending only on a spacetime metric and 3-form may be dual to a theory that has a trivector which cannot be gauged away. The four-dimensional case will be covered in section \ref{Sd4}. Section \ref{SExamples} contains explicit examples of the dualities acting on M2-branes. We find that when the duality transformation acts along the worldvolume of the brane, it acts like the Harrison transformation, charging solutions. The difficulties found in the examples in \cite{Malek:2012pw} are now removed, forcing us instead to describe the dual solutions using a trivector for those cases. We also act with transformations of $SL(5)$ on smeared extreme M2-branes and find that the dual solutions carry momentum in the transverse direction that the brane was delocalised along. Finally, we will discuss our results and justify our interpretation of timelike dualities, which is based on an analogy with geometry, in section \ref{SConclusions}.

\section{Generalised geometry} \label{SGenGeom}
In this section, we will give a brief overview of generalised geometry and how it can be used to make the U-duality symmetry of 11-dimensional supergravity manifest. The aim is to use it to find objects which transform as tensors under U-duality. We begin by looking at the coordinates and will then see how to combine the bosonic fields.\footnote{We ignore fermions throughout.}

In string theory, T-duality exchanges momenta and string winding numbers
\begin{equation}
 P_i \leftrightarrow W^i \, ,
\end{equation}
and, similarly in M-theory, U-duality mixes momenta and membrane wrapping modes. We restrict the dualities to act only along $d<5$ directions, forming the $E_d$ duality group, as listed in table \ref{TDualityGroups} so that we only need to take into account the wrapping modes due to the M2-brane. The other spacetime directions form a transverse undualisable spacetime and we will impose certain requirements on the bosonic fields as we will shortly explain.\footnote{Although we refer to the ``dualisable'' spacetime and transverse spacetime, only one of these will include time and will be a space\emph{time} while the other is just a ``space''. We do this because at this stage we want to keep the discussion general and thus do not specify whether time is dualised or not.} The M2-brane winding modes are labelled by an antisymmetric 2-tensor $Z^{ij}$ so that U-dualities mix
\begin{equation}
 P_i \leftrightarrow Z^{ij} \, ,
\end{equation}
where the indices $i, j = 1, \ldots d$ label the dualisable spacetime directions.\footnote{Readers familiar with 11-dimensional supergravity will recognize this as a central charge of the supersymmetry algebra.} Just as the momenta are conjugate to spacetime directions
\begin{equation}
 x^i = \frac{\delta}{\delta P_i} \, ,
\end{equation}
\begin{table}
\centering
\begin{tabular}{|c||c|c||c|}
\hline
d & $E_d$ & $H_d$ & $\tilde{H}_d$ \\
\hline
3 & $SL(3) \times SL(2)$ & $SO(3) \times SO(2)$ & $SO(2,1) \times SO(1,1)$ \\
4 & $SL(5)$ & $SO(5)$ & $SO(3,2)$ \\
5 & $SO(5,5)$ & $SO(5) \times SO(5)$ & $SO(5,C)$ \\
6 & $E_6$ & $USp(8)$ & $USp(4,4)$ \\
7 & $E_7$ & $SU(8)$ & $SU^*(8)$ \\
8 & $E_8$ & $SO(16)$ & $SO^*(16)$ \\
\hline
\end{tabular}
\caption{The U-duality groups $E_d$, their maximal compact subgroups $H_d$ and the non-compact subgroups that arise in timelike dualities $\tilde{H}_d$ \cite{Hull:1998br}.}
\label{TDualityGroups}
\end{table}
where the derivatives are understood in the usual sense as acting on momentum eigenstates, we can introduce ``dual'' coordinates $y_{ij}$, which are conjugate to these wrapping modes
\begin{equation}
 y_{ij} = \frac{\delta}{\delta Z^{ij}} \, .
\end{equation}
It is important to note that the $y_{ij}$ are antisymmetric and will, together with the dualisable spacetime coordinates $x^i$, form a representation space of the $E_d$ duality group. We call these the ``generalised coordinates'',
\begin{equation}
 X^M = \left( \begin{array}{c}
              x^i \\
              \frac{1}{\sqrt{2}} y_{ij} \\
              x^A
              \end{array} \right) \, ,
\end{equation}
where we have also included the transverse spacetime coordinates, labelled by the index $A = d+1, \ldots, 11$. For example, when $d=3$, the three dualisable spacetime coordinates and their three duals belong to the 6-dimensional representation of $SL(2)\times SL(3)$, while for $d=4$ they transform under the $10=4+6$-dimensional representation of $SL(5)$. We will see explicitly how they transform under U-dualities in section \ref{SEDuality}. The transverse spacetime coordinates $x^A$ transform as $\bar{d}$-vectors under $GL(\bar{d})$, where $\bar{d} = 11 - d$. This is the group of rigid diffeomorphisms acting on the transverse spacetime.

Similarly, U-duality mixes the metric and 3-form and so in order to make the action of U-duality manifest we combine them into a generalised metric. For four-dimensional dualities this was originally found by studying dualities on the membrane worldvolume \cite{Duff:1990hn} and has more recently been constructed as a non-linear realisation of $E_d \times GL(\bar{d})$ for duality groups in $d<8$ dimensions \cite{Berman:2011jh,Malek:2012pw}. The assumption is made that the spacetime metric is factorisable so that it has no mixed components along the dualisable and transverse undualisable spacetime and we can write its components as $g_{ab} = \left(g_{ij}, g_{AB}\right)$, where the indices $a,b = 1, \ldots 11$ label all eleven directions, while $i, j = 1, \ldots d$ label the dualisable directions and $A, B = d + 1, \ldots 11$ label the transverse spacetime and there are no mixed components $g_{iA}$. Similarly the 3-form $C_3$ is taken to only have non-zero components along the dualisable space, $C_{ijk}$. For $d = 3, 4$ the result is similar to \cite{Duff:1990hn,Hull:2007zu} but differs by a conformal factor:
\begin{equation}
 \gm_{MN} = |g_{11}|^{-1/2} \left( \begin{array}{ccc}
                               g_{ij} + \frac{1}{2} C_{imn} C^{mn}_{\ph{mn}j} & \frac{1}{\sqrt{2}} C_i^{\ph{i}kl} & 0 \\
				\frac{1}{\sqrt{2}} C^{ij}_{\ph{ij}k} & g^{i[k} g^{l]j} & 0 \\
				0 & 0 & g_{AB}
                              \end{array} \right) \, , \label{EEGM}
\end{equation}
where $|g_{11}|$ is the determinant of the 11-dimensional metric. We will often drop the indices and write this as
\begin{equation}
 \gm = |g_{11}|^{-1/2} \left( \begin{array}{ccc}
           g + \frac{1}{2} C g^{-1} g^{-1} C & \frac{1}{\sqrt{2}} C g^{-1} & 0 \\
           \frac{1}{\sqrt{2}} g^{-1} C & g^{-1} g^{-1} & 0 \\
           0 & 0 & g_{\bar{d}}
           \end{array} \right) \, ,
\end{equation}
where $g$ without a subscript will always be taken to signify the components along the dualisable directions. The conformal factor is crucial as otherwise the generalised metric does not transform correctly under U-dualities, as shown in section \ref{SEDuality}.

It is useful to extend the definition of a U-duality from an element of $E_d$ to an element of $\mathcal{E}_d \equiv E_d \times GL(\bar{d})$ so that for $U_e \in \mathcal{E}_d$ the generalised metric transforms as
\begin{equation}
 \gm \rightarrow \left(U_e\right)^T \gm U_e \, .
\end{equation}
The part of $U_e$ belonging to $GL(\bar{d})$ is trivial and we will often ignore it.\footnote{We will use the phrase ``trivial'' throughout this paper for dualities that only act as gauge transformations.} Correspondingly we write the generalised metric as
\begin{equation}
 \gm = |g_{11}|^{-1/2} \left( \begin{array}{cc}
           g + \frac{1}{2} C g^{-1} g^{-1} C & \frac{1}{\sqrt{2}} C g^{-1} \\
           \frac{1}{\sqrt{2}} g^{-1} C & g^{-1} g^{-1} \\
           \end{array} \right) \, .
\end{equation}
Because the generalised metric transforms naturally under U-duality, we view it as the fundamental physical variable describing the theory. One thus expects that the low-energy effective action can be written in terms of $\gm$ directly, rather than $g_{11}$ and $C_3$ separately. \footnote{For fermions one would have to use the generalised vielbein instead.}

Under a U-duality, the generalised coordinates transform contravariantly
\begin{equation}
 X \rightarrow \left(U_e\right)^{-1} X \,. \label{ECT}
\end{equation}
We define generalised derivatives corresponding to the generalised coordinates by
\begin{equation}
 \begin{split}
  \partial_M &\equiv \frac{\partial}{\partial X^M} \\
  &= \left( \begin{array}{c}
             \partial_i \\
	     \frac{1}{\sqrt{2}} \partial^{ij} \\
	     \partial_A
            \end{array} \right)\,.
 \end{split}
\end{equation}
Using the generalised metric and coordinates one can write a manifestly duality invariant Lagrangian, here given for $d=4$.
\begin{equation}
 \begin{split}
 \mathcal{L} &= \frac{1}{12} \gm^{MN} \partial_M \gm^{PQ} \partial_N \gm_{PQ} - \frac{1}{2} \gm^{MN} \partial_M \gm^{PQ} \partial_P \gm_{NQ} \\
 & \quad + \frac{1}{108} \gm^{MN} \left( \gm^{KL} \partial_M \gm_{KL} \right) \left( \gm^{PQ} \partial_N \gm_{PQ} \right) \\
 & \quad + \frac{1}{6} \gm^{MN} \partial_M \left( |g_{11}|^{1/2} g^{AB} \right) \partial_N \left( |g_{11}|^{-1/2} g_{AB} \right)\,. \label{ELEEA}
 \end{split}
\end{equation}
By using the solution to the section condition\footnote{See
\cite{Berman:2011cg} for a group-invariant section condition for $E_4 = SL(5)$
and \cite{Berman:2012vc} for the U-duality groups $E_5 \ldots E_8$.}
\begin{equation}
 \partial^{ij} \gm = 0\,,
\end{equation}
such that all fields depend only on the spacetime coordinates and not the dual $y_{ij}$, the Lagrangian reduces to the conventional one
\begin{equation}
 \mathcal{L} = \sqrt{|g_{11}|} \left( R - \frac{1}{48} F^2 \right)\,,
\end{equation}
up to a boundary term that can also be put in a U-duality invariant form \cite{Berman:2011kg}. Here $R$ is the 11-dimensional Ricci scalar and $F = dC_3$ is the four-form field strength associated to $C_3$.\footnote{The Chern-Simons term vanishes here because the 3-form has non-vanishing components only in the dualisable directions.}

\subsection{Generalised vielbeins} \label{SGenVielbein}
The generalised metric parameterises the coset
\begin{equation}
 \frac{E_d \times GL(\bar{d})}{H_d \times SO(\bar{d}-1,1)}\,,
\end{equation}
where $H_d$ is the maximal compact subgroup of $E_d$, as given in table \ref{TDualityGroups}. The coset
\begin{equation}
 \frac{E_d}{H_d}
\end{equation}
is parameterised by the bosonic fields along the dualisable directions, $g$ and
$C_3$, while the second factor, the coset
\begin{equation}
 \frac{GL(\bar{d})}{SO(\bar{d}-1,1)}\,,
\end{equation}
is parameterised by the Lorentzian spacetime metric in the transverse space,
$g_{\bar{d}}$. The maximal compact subgroup $H_d$ acts as a local symmetry group
and its action can be made explicit by decomposing the generalised metric in
terms of a generalised vielbein
\begin{equation}
 \gm = L^T \eta_E L\,,
\end{equation}
where the generalised flat line element is
\begin{align}
 dS^2 &= dX^T \eta_E dX \nonumber \\
 &= \sum_i dx^i dx^i + \frac{1}{2} \sum_{i,j} dy_{ij} dy_{ij} + \eta_{AB} dx^A dx^B\,.
\end{align}
Here $\eta_{AB}$ is the $\bar{d}$-dimensional Minkowski metric of the transverse space. An element of the extended U-duality group $\mathcal{E}_d$ acts on the generalised vielbein through a right-action
\begin{equation}
 L \rightarrow L U_e\,,
\end{equation}
while an element of the ``extended'' local symmetry group, $h \in H_d \times SO(\bar{d}-1,1)$, acts through a left-action
\begin{equation}
 L \rightarrow h L\,.
\end{equation}
We see that the local symmetry group $H_d \times SO(\bar{d}-1,1)$ is the group of transformations preserving the internal metric $\eta_E$.

The generalised vielbein can chosen to be lower-triangular, given by
\begin{equation}
 L^{\bar{M}}_{\ph{M}N} = |\tilde{e}_{11}|^{-1/2} \left( \begin{array}{ccc}
                        \tilde{e}^{\bar{j}}_{\ph{j}i} & 0 & 0 \\
                        \frac{1}{\sqrt{2}} C_{\bar{i}\bar{j}k} & e^{\ph{[\bar{k}}[i}_{[\bar{k}} e^{\ph{\bar{l}]}j]}_{\bar{l}]} & 0 \\
                        0 & 0 & \tilde{e}^{\bar{A}}_{\ph{A}B}
                        \end{array} \right)\,,
\end{equation}
where $\bar{M}$ labels the generalised flat tangent space coordinates $X^{\bar{M}} = \left(x^{\bar{i}}, \frac{1}{\sqrt{2}} y_{\bar{i}\bar{j}}, x^{\bar{A}}\right)$. We will write this without explicit indices as
\begin{equation}
 L = |\tilde{e}_{11}|^{-1/2} \left( \begin{array}{ccc}
                        \tilde{e} & 0 & 0 \\
                        \frac{1}{\sqrt{2}} e e C & e e & 0 \\
                        0 & 0 & \tilde{e}_{\bar{d}}
                        \end{array} \right) \label{EGV}\,.
\end{equation}
We will now drop the components along the transverse space for simplicity. One could equally well have chosen an upper triangular vielbein
\begin{equation}
 L_\Omega = |\tilde{\bar{e}}_{11}|^{-1/2}
                    \left( \begin{array}{cc}
                    \tilde{\bar{e}} & \frac{1}{\sqrt{2}} \tilde{\bar{e}} \Omega \\
                    0 & \bar{e} \bar{e}
                    \end{array} \right)\,,
\end{equation}
where $\Omega^{ijk}$ is a trivector, totally antisymmetric in its indices. The generalised metric would then be written as \footnote{See \cite{Andriot:2011uh} for a detailed discussion of this change of variables as used in the $O(d,d)$ case.}
\begin{equation}
 \gm = |\og_{11}|^{-1/2} \left( \begin{array}{cc}
           \og & \frac{1}{\sqrt{2}} \og \Omega \\
           \frac{1}{\sqrt{2}} \Omega \og & \og^{-1} \og^{-1} + \frac{1}{2} \Omega \og \Omega \\
           \end{array} \right) \label{EGMO} \, .
\end{equation}
We will focus on the $d=3$ case for most of this paper as it allows us to reach the physically significant conclusions without the extra complication of more dimensions. We refer the reader to \cite{Malek:2012pw} for details on $d=4$ equations in the Euclidean case and section \ref{Sd4} for the timelike case. We begin by defining the dualised 3-form and tri-vector
\begin{equation}
 \begin{array}{cc}
  V = \frac{1}{3!} \epsilon^{ijk} C_{ijk} \, ,\quad  & W = \frac{1}{3!} \bar{\epsilon}_{ijk} \Omega^{ijk} \, ,
  \end{array}
\end{equation}
where $\epsilon^{ijk}$ is the Levi-Civita tensor in the three dimensions to be dualised defined with respect to $g$ while $\bar{\epsilon}_{ijk}$ is the three-dimensional Levi-Civita tensor with respect to $\og$. In terms of these objects we can for $d=3$ identify
\begin{equation} \label{3CtoOmega}
 \begin{split}
 \og_{ij} &= g_{ij} \left(1 + V^2\right)^{2/3} \, , \\
 \Omega^{ijk} &= \frac{\epsilon^{ijk}V}{1+V^2} = \frac{g^{im} g^{jn} g^{ko} C_{mno}}{1+V^2}\, , \\
 \og_{AB} &= g_{AB} \left(1 + V^2\right)^{-1/3}
 \end{split}
\end{equation}
and inversely
\begin{equation} \label{3OmegatoC}
 \begin{split}
 g_{ij} &= \og_{ij} \left(1+W^2\right)^{-2/3} \, , \\
 C_{ijk} &= \frac{\bar{\epsilon}_{ijk}W}{1+W^2} = \frac{\og_{im} \og_{jn} \og_{ko} \Omega^{mno}}{1+W^2} \, , \\
 g_{AB} &= \og_{AB} \left(1 + W^2\right)^{1/3} \, .
 \end{split}
\end{equation}

In the Euclidean case one can always choose to describe the generalised metric in terms of the fields $\left(g_{11}, C_3\right)$ or $\left(\og_{11}, \Omega_3\right)$ or a combination $\left(\hg_{11}, C_3, \Omega_3\right)$. This is a choice of frame or a choice of ``preferred fields'' in the language of non-linear realisations and if we view the generalised metric as the fundamental variable, they are both equally valid. The choice to use the $\left(g_{11}, C_3\right)$ frame can be seen as simply a convention. We can explicitly show that the choice of frame is arbitrary because we can always rotate an upper triangular vielbein into a lower triangular one
\begin{equation}
 L_C = H L_\Omega \, ,
\end{equation}
where $H \in SO(2)$ is given by
\begin{equation}
 H = \left( \begin{array}{cc}
            \cos \theta \delta^{\bar{i}}_{\ph{\bar{i}}\bar{k}} & \frac{1}{\sqrt{2}} \sin \theta \epsilon^{\bar{i}\bar{k}\bar{l}} \\
            \frac{1}{\sqrt{2}} \sin \theta \epsilon_{\bar{i}\bar{j} \bar{k}} & \cos \theta \delta_{\bar{i}\bar{j}}^{\ph{\bar{i}\bar{j}}\bar{k}\bar{l}}
            \end{array} \right)\,,
\end{equation}
and the trivector is gauged away when choosing
\begin{equation}
 \tan\theta = W\,.
\end{equation}
Here $\epsilon_{\bar{1}\bar{2}\bar{3}} = - 1$ is the totally antisymmetric tensor in the tangent spacetime. Thus, the trivector $\Omega_3$ can always be gauged away. However, we will see in section \ref{STIntSym} this is not generally the case when time is included. A similar issue in the $O(d,d)$ case is discussed in \cite{Andriot:2011uh}.

\subsection{Spacelike dualities} \label{SEDuality}
Before we move on to include time amongst the dualisable coordinates, we will quickly review the action of dualities in the Euclidean case. More details can be found in \cite{Malek:2012pw}.

We can decompose the U-duality group $E_d$ into its ``geometric'' $SL(d)$ subgroup
\begin{equation}
 U_{SL(d)} = \left( \begin{array}{cc}
                    A & 0 \\
                    0 & A^{-T} A^{-T}
                    \end{array} \right)\,, \label{ESL(d)}
\end{equation}
which mixes the dualisable directions and their duals amongst themselves:
\begin{equation}
 \begin{split}
  x^i &\rightarrow A^i_{\ph{i}j} x^j\,, \\
  y_{ij} & \rightarrow \left(A^{-1}\right)_{i}^{\ph{i}k} \left(A^{-1}\right)_{j}^{\ph{j}l} y_{kl}\,.
 \end{split}
\end{equation}

The quotient group $E_d/SL(d)$ can be split into $\frac{d!}{3!(d-3)!}$ non-commuting $SL(2)$ subgroups, one for each set of three dualisable directions. Each of these $SL(2)$ subgroups contains the three elements
\begin{align}
 \textrm{dilatations, } U_{\alpha} &= \left( \begin{array}{cc}
                      \alpha^{-1} & 0 \\
                      0 & \alpha
                      \end{array} \right)\,, \label{ESL(2)1} \\
 \textrm{C-shifts, } U_C &= \left( \begin{array}{cc}
                1 & 0 \\
                \frac{1}{\sqrt{2}} C & 1
                \end{array} \right)\,, \label{ESL{2}2}\\
 \textrm{$\Omega$-shifts, } U_{\Omega} &= \left( \begin{array}{cc}
                      1 & \frac{1}{\sqrt{2}} \Omega \\
                      0 & 1
                      \end{array} \right)\,, \label{ESL(2)3}
\end{align}
where $C$ and $\Omega$ have only one non-zero component along the three directions to which the $SL(2)$ belongs. For $d=3$ this is particularly simple because there is only one such $SL(2)$ subgroup as the duality group is $E_3 = SL(3) \times SL(2)$. For $d=4$ the duality group $E_4 = SL(5)$ contains the geometric $SL(4)$ subgroup and three $SL(2)$ subgroups as outlined above.

The $U_\alpha$ acts by dilatations $g \rightarrow g \alpha^{-1}$ while the $C$-shifts and $\Omega$-shifts shift the 3-form $C_3$ and trivector $\Omega_3$, respectively. Thus each of these last two transformations is trivial in some frame but in the $\left(g_{11}, C_3\right)$ frame the $\Omega$-shift is non-trivial, while the $C$-shift is non-trivial in the $\left(\og_{11},\Omega_3\right)$ frame. In the $\left(g_{11}, C_3\right)$, the action of the $U_\Omega$ shift for $d=3$ is given by
\begin{equation}
 \begin{split} \label{3OmegaDuality}
 g'_{ij} &= g_{ij} \left((1+AC_{123})^2 + A^2 |g_3| \right)^{-2/3}\,, \\
 g'_{AB} &= g_{AB} \left((1+AC_{123})^2 + A^2 |g_3| \right)^{1/3}\,, \\
 C'_{123} &= \frac{C_{123} \left(1 + A C_{123}\right) + A|g_3|}{\left(1+AC_{123}\right)^2 + A^2 |g_3|}\,.
 \end{split}
\end{equation}
One can also construct a Buscher duality \cite{Buscher:1987sk,Buscher:1987qj} by performing three successive transformations
\begin{equation}
 \begin{split}
  U_B &= U_C U_\Omega U_C \\
      &= \left( \begin{array}{cc}
            0 & \frac{1}{\sqrt{2}} \Omega \\
            \frac{1}{\sqrt{2}} C & 0
            \end{array} \right)\,, \label{EBuscherGenerated}
 \end{split}
\end{equation}
where
\begin{equation}
 \begin{split}
 \Omega^{123} &= A \,,\\
 C_{123} &= - 1/A\,.
 \end{split}
\end{equation}
The transformed fields are
\begin{equation}
 \begin{split} \label{3BuscherDuality}
  g'_{ij} &= g_{ij} \left( A^2 \left( C_{123}^2 + |g_3| \right) \right)^{-2/3}\,, \\
  g'_{AB} &= g_{AB} \left( A^2 \left( C_{123}^2 + |g_3| \right) \right)^{1/3}\,, \\
  C'_{123} &= - \frac{C_{123}}{A^2 \left( C_{123}^2 + |g_3| \right)}\,.
 \end{split}
\end{equation}

Under this $SL(2)$, the generalised coordinates are split into three pairs, mixing the spacetime and dual coordinates
\begin{equation}
 \left( \begin{array}{c} x^1 \\ y_{23} \end{array} \right), \left( \begin{array}{c} x^2 \\ -y_{13} \end{array} \right), \left( \begin{array}{c} x^3 \\ y_{12} \end{array} \right)\,.
\end{equation}
$U_C$ rotates the pairs one way (for $C_{123} = A$)
\begin{equation}
 \begin{split}
 \left( \begin{array}{c} x^1 \\ y_{23} \end{array} \right) & \rightarrow \left( \begin{array}{c} x^1 \\ y_{23} - A x^1 \end{array} \right)\,, \\
 \left( \begin{array}{c} x^2 \\ -y_{13} \end{array} \right) & \rightarrow \left( \begin{array}{c} x^2 \\ -y_{13} - A x^2 \end{array} \right)\,, \\
 \left( \begin{array}{c} x^3 \\ y_{12} \end{array} \right) & \rightarrow \left( \begin{array}{c} x^3 \\ y_{12} - A x^3 \end{array} \right)\,,
 \end{split}
\end{equation}
while $U_\Omega$ rotates them the other way (for $\Omega^{123} = A$)
\begin{equation}
 \begin{split}
 \left( \begin{array}{c} x^1 \\ y_{23} \end{array} \right) & \rightarrow \left( \begin{array}{c} x^1 - A y_{23} \\ y_{23} \end{array} \right)\,, \\
 \left( \begin{array}{c} x^2 \\ -y_{13} \end{array} \right) & \rightarrow \left( \begin{array}{c} x^2 + A y_{13} \\ -y_{13} \end{array} \right)\,, \\
 \left( \begin{array}{c} x^3 \\ y_{12} \end{array} \right) & \rightarrow \left( \begin{array}{c} x^3 - A y_{12} \\ y_{12} \end{array} \right)\,,
 \end{split}
\end{equation}
and $U_\alpha$ acts on the doublets as
\begin{equation}
 \begin{split}
 \left( \begin{array}{c} x^1 \\ y_{23} \end{array} \right) & \rightarrow \left( \begin{array}{c} \alpha x^1 \\ \alpha^{-1} y_{23} \end{array} \right)\,, \\
 \left( \begin{array}{c} x^2 \\ -y_{13} \end{array} \right) & \rightarrow \left( \begin{array}{c} \alpha x^2 \\ - \alpha^{-1} y_{13} \end{array} \right)\,, \\
 \left( \begin{array}{c} x^3 \\ y_{12} \end{array} \right) & \rightarrow \left( \begin{array}{c} \alpha x^3 \\ \alpha^{-1} y_{12} \end{array} \right)\,.
 \end{split}
\end{equation}

\section{Lorentzian generalised metric} \label{STGenGeom}
We now include time amongst the three dualisable directions and construct the generalised metric in a similar fashion but using a different generalised flat line element. This generalised flat line element will be preserved by the non-compact subgroups $\tilde{H}_d$ listed in table \ref{TDualityGroups}. The generalised metric then parameterises the ``Lorentzian'' coset space
\begin{equation}
 \frac{E_d \times GL(\bar{d})}{\tilde{H}_d \times SO(\bar{d})}\,.
\end{equation}

\subsection{Lorentzian coset space for $d=3$}
The U-duality group  in the Lorentzian case is still $E_d$ as is to be expected by analogy with geometry. The $d$-dimensional metric parameterises the coset $\frac{GL(d)}{SO(d)}$ in the Euclidean and $\frac{GL(d)}{SO(d-1,1)}$ in the Lorentzian case. Clearly it is the local symmetry group $SO(d)$ vs. $SO(d-1,1)$ which contains the information about the signature of the spacetime. Similarly for U-duality we find that the local symmetry group $H_d \times SO(\bar{d}-1,1)$ changes to $\tilde{H}_d \times SO(\bar{d})$ with $H_d$ and $\tilde{H}_d$ given in table \ref{TDualityGroups}.

To construct the generalised flat line element we want to interpret the action of $\tilde{H}_3 = SO(1,1) \times SO(2,1)$ on the generalised coordinates. To do so, we compare it to the action of the U-duality group $SL(2) \times SL(3)$ where $SO(2,1) \subset SL(3)$ and $SO(1,1) \subset SL(2)$. The $SL(3)$ interchanges the three dualisable spacetime indices amongst each other and so we interpret its non-compact subgroup $SO(2,1)$ as the local Lorentz group for the three dualisable spacetime indices. As expected for the three dualisable spacetime directions the flat line element then has to be Lorentzian
\begin{equation}
 ds^2 = - dt^2 + \sum_\mu dx^\mu dx^\mu\,.
\end{equation}
Here the indices $\mu = 2, 3$ run over the spatial indices so that
\begin{equation}
 x^{i} = \left( t, x^{\mu} \right)\,.
\end{equation}

We have seen in section \ref{SEDuality} that the $SL(2)$ subgroup causes rotations within each of the three doublets, here taking $x^1 \rightarrow t$
\begin{equation}
 \left( \begin{array}{c} t \\ y_{23} \end{array} \right), \left( \begin{array}{c} x^2 \\ -y_{t3} \end{array} \right), \left( \begin{array}{c} x^3 \\ y_{t2} \end{array} \right)\,.
\end{equation}
Its local symmetry group is $SO(1,1)$ so that we associate a Lorentzian metric
\begin{equation}
 \eta_{SO(1,1)} = \left( \begin{array}{cc}
                  -1 & 0 \\
		  0 & 1
                 \end{array} \right)
\end{equation}
with each doublet. This then gives the generalised flat line element as
\begin{equation}
 dS^2 = - dt^2 + \sum_{\mu} dx^\mu dx^\mu - \sum_{\mu} dy_{t\mu} dy_{t\mu} + \frac{1}{2} \sum dy_{\mu\nu} dy_{\mu\nu} + \sum_A dx^A dx^A \label{ETA3}\,.
\end{equation}
It is evident that the dual coordinates $y_{\mu\nu}$ are spacelike, while the $y_{t\mu}$ are timelike.

We can now construct the generalised metric as
\begin{equation}
 \gm = L_C^T \mathcal{M} L_C\,,
\end{equation}
where $\mathcal{M}$ is the generalised internal metric giving the generalised flat line element \eqref{ETA3} and $L_C$ is the generalised vielbein \eqref{EGV}. This gives a generalised metric of the same form as before, equation \eqref{GMC},
\begin{equation}
 \gm = |g_{11}|^{-1/2} \left( \begin{array}{ccc}
                                g + \frac{1}{2} C g^{-1} g^{-1} C & \frac{1}{\sqrt{2}} C g^{-1} g^{-1} & 0 \\
                                \frac{1}{\sqrt{2}} g^{-1} g^{-1} C & g^{-1} g^{-1} & 0 \\
                                0 & 0 & g_{8} \end{array} \right)\,, \label{GMT3}
\end{equation}
where now $g$ is Lorentzian and $g_{8}$ is Euclidean. The duality invariant action \eqref{ELEEA} can now be expressed in terms of this ``Lorentzian'' generalised metric to include time.

\subsection{Lorentzian coset space for $d=4$}
We now wish to construct the generalised metric parameterising the coset
\begin{equation}
 \frac{SL(5) \times GL(7)}{SO(3,2) \times SO(7)}\,.
\end{equation}
We begin by finding the generalised flat line element preserved by the local symmetry group $SO(3,2) \times SO(6)$. The latter factor is the local rotations group  of the transverse undualisable space. To understand how the first factor acts on the generalised coordinates we first write the dualisable spacetime coordinates and their duals $\left(x^i, y_{ij}\right)$ in terms of $SL(5)$ covariant indices \cite{Berman:2011cg}. The 10 coordinates belong to the antisymmetric representation of $SL(5)$
\begin{equation}
 X^{\hat{m}\hat{n}} = \left\{ \begin{array}{rl}
                  X^{i5} &= x^i \,,\\
                  X^{ij} &= \frac{1}{2} \epsilon^{ijkl} y_{kl}\,,
                  \end{array} \right.
\end{equation}
where $\hat{m}, \hat{n} = 1, \ldots 5$ are $SL(5)$ indices and $\epsilon^{ijkl}$ is the Levi-Civita tensor for the flat dualisable spacetime. Because the spacetime coordinates $x^i$ have one timelike and four spacelike directions we associate the second timelike direction with the $5$ index. We write
\begin{equation}
 \eta_{SO(3,2)} = \left( \begin{array}{ccccc}
                        -1 & 0 & 0 & 0 & 0 \\
			0 & 1 & 0 & 0 & 0 \\
			0 & 0 & 1 & 0 & 0 \\
			0 & 0 & 0 & 1 & 0 \\
			0 & 0 & 0 & 0 & -1
                       \end{array} \right)\,,
\end{equation}
so the generalised flat line element is
\begin{equation}
 \begin{split}
  dS^2 &= -dX^{\hat{m}\hat{n}} dX^{\hat{p}\hat{q}} \eta_{\hat{m}\hat{p}} \eta_{\hat{n}\hat{q}} + \sum_A dx^A dx^A \\
  &= -dt^2 + \sum_\mu dx^{\mu} dx^{\mu} - \sum_{\mu} dy_{t\mu} dy_{t\mu} + \frac{1}{2} \sum_{\mu,\nu} dy_{\mu\nu} dy_{\mu\nu} + \sum_A dx^A dx^A\,, \label{ETA4}
 \end{split}
\end{equation}
where $\mu, \nu = 2, 3, 4$ once again run over the spatial indices so that $x^i = \left(t, x^\mu\right)$. We are using the minus sign in $dS^2 = - dX^{\hat{m}\hat{n}} dX^{\hat{p}\hat{q}} \eta_{\hat{m}\hat{p}} \eta_{\hat{n}\hat{q}}$ to obtain a flat line element that is mostly positive along the spacetime components in order to keep our conventions consistent. Once again we find that the dual coordinates $y_{ij}$ are spacelike for $y_{\mu\nu}$ and timelike for $y_{t\mu}$. 

The generalised metric takes a similar form to the three-dimensional case
\begin{equation}
 \begin{split}
  \gm &= L_C^T \mathcal{M} L \\
  &= |g_{11}|^{-1/2} \left( \begin{array}{ccc}
                                g + \frac{1}{2} C g^{-1} g^{-1} C & \frac{1}{\sqrt{2}} C g^{-1} g^{-1} & 0 \\
                                \frac{1}{\sqrt{2}} g^{-1} g^{-1} C & g^{-1} g^{-1} & 0 \\
                                0 & 0 & g_{7} \end{array} \right)\,.
 \end{split}
\end{equation}

\subsection{Internal rotations, the 3-form and the trivector} \label{STIntSym}
We will now focus on the case where the duality group acts in three directions, including time. This example carries all the relevant physics but does not have the complication coming from having four dimensions and thus possibly several independent non-zero components of the 3-form and trivector. We will give the relevant formulae for the $d=4$ case in section \ref{Sd4}.

The generalised vielbeins transform under the local symmetry group, $H_3$ in the Euclidean and $\tilde{H}_3$ in the Lorentzian case. We have so far given it in lower-triangular form
\begin{equation}
 L_C = |\tilde{e}_{11}|^{-1/2} \left( \begin{array}{ccc}
                        \tilde{e} & 0 & 0 \\
                        \frac{1}{\sqrt{2}} e e C & e e & 0 \\
                        0 & 0 & \tilde{e}_{8}
                        \end{array} \right)\,.
\end{equation}
In the Euclidean case, we have shown that one can always chose this parameterisation. The caveat is for non-geometric backgrounds where topological obstructions hinder the local rotation needed to remove the trivector field \cite{Grana:2008yw,Andriot:2011uh,Andriot:2012wx,Andriot:2012an,Aldazabal:2011nj,Dibitetto:2012rk}. Instead the trivector field is shown in these works to give rise to non-geometric fluxes. Barring such obstructions the trivector field can always be gauged away to obtain a supergravity solution containing a metric and 3-form.

For timelike dualities we also encounter an obstruction. Now the local symmetry group excluding Lorentz transformations, $\tilde{H}_3 = SO(1,1)$, contains hyperbolic rotations. Starting with an upper triangular vielbein $L_\Omega$ we can rotate it into a lower triangular one $L_C = H L_\Omega$ by
\begin{equation}
 H = \left( \begin{array}{cc}
            \cosh \theta \delta^{\bar{i}}_{\ph{\bar{i}}\bar{k}} & \frac{1}{\sqrt{2}} \sinh \theta \epsilon^{\bar{i}\bar{k}\bar{l}} \\
            -\frac{1}{\sqrt{2}} \sinh \theta \epsilon_{\bar{i}\bar{j} \bar{k}} & \cosh \theta \delta_{\bar{i}\bar{j}}^{\ph{\bar{i}\bar{j}}\bar{k}\bar{l}}
            \end{array} \right)\,, \label{EHInt}
\end{equation}
when choosing
\begin{equation}
 \tanh^2 \theta = W^2\,.
\end{equation}
However, because $\tanh^2 \theta < 1$ this choice is only possible when
\begin{equation}
 W^2 < 1\,. \label{ELTU1}
\end{equation}
We find that if $W^2 < 1$ one can change the frame from the $\left(\og_{11}, \Omega_3\right)$ to the $\left(g_{11}, C_3\right)$ frame, finding
\begin{equation}
 \begin{split}
 g_{ij} &= \og_{ij} \left(1-W^2\right)^{-2/3}\,, \\
 C_{ijk} &= \frac{\bar{\epsilon}_{ijk}W}{1-W^2} = \frac{\og_{im} \og_{jn} \og_{ko} \Omega^{mno}}{1-W^2}\,, \\
 g_{AB} &= \og_{AB} \left(1 - W^2\right)^{1/3}\,.
 \end{split} \label{ELTU2}
\end{equation}
Similarly, if $V^2 < 1$ one can change from the $\left(g_{11}, C_3\right)$ to the $\left( \og_{11}, \Omega_3\right)$ frame by the inverse transformation
\begin{equation} \label{3CtoOmegaD}
 \begin{split}
 \og_{ij} &= g_{ij} \left(1 - V^2\right)^{2/3}\,, \\
 \Omega^{ijk} &= \frac{\epsilon^{ijk}V}{1-V^2} = \frac{g^{im} g^{jn} g^{ko} C_{mno}}{1-V^2}\,, \\
 \og_{AB} &= g_{AB} \left(1 - V^2\right)^{-1/3}\,.
 \end{split}
\end{equation}
We see that there may be situations where one has to consider a non-zero trivector field that cannot be gauged away because $W^2 \geq 1$. We will construct explicit examples by the use of timelike dualities in section \ref{SExamples}.

We now briefly pause to make an analogy with geometry. There one describes the
system through a metric which we take to be two-dimensional for simplicity.
\begin{equation}
 g = \left( \begin{array}{cc}
            g_{11} & g_{12} \\
            g_{12} & g_{22}
            \end{array} \right)\,.
\end{equation}
This is symmetric and parameterises the coset $\frac{GL(2)}{SO(2)}$ in the Euclidean case. It thus can be expressed in terms of a vielbein
\begin{equation}
 e^{\bar{a}}_{\ph{\bar{a}}b} = \left( \begin{array}{cc}
                                      e^{\bar{1}}_{\ph{\bar{1}}1} & e^{\bar{1}}_{\ph{\bar{1}}2} \\
                                      e^{\bar{2}}_{\ph{\bar{1}}1} & e^{\bar{2}}_{\ph{\bar{1}}2}
                                      \end{array} \right)\,,
\end{equation}
but this can be simplified by local $SO(2)$ rotations to give a vielbein in upper triangular or lower triangular form
\begin{align}
 \tilde{e} &= \left( \begin{array}{cc}
                    \tilde{e}^{\bar{1}}_{\ph{\bar{1}}1} & 0 \\
                    \tilde{e}^{\bar{2}}_{\ph{\bar{1}}1} & \tilde{e}^{\bar{2}}_{\ph{\bar{1}}2}
                    \end{array} \right)\,, \\
 \hat{e} &= \left( \begin{array}{cc}
                    \hat{e}^{\bar{1}}_{\ph{\bar{1}}1} & \hat{e}^{\bar{1}}_{\ph{\bar{1}}2} \\
                    0 & \hat{e}^{\bar{2}}_{\ph{\bar{1}}2}
                    \end{array} \right)\,.
\end{align}
The metric can be expressed in terms of these two vielbeins as
\begin{equation}
 \begin{split} \label{EEGL}
 g_{11} &= \left(\tilde{e}^{\bar{1}}_{\ph{\bar{1}}1}\right)^2 + \left(\tilde{e}^{\bar{2}}_{\ph{\bar{1}}1}\right)^2\,, \\
 g_{12} &= \tilde{e}^{\bar{2}}_{\ph{\bar{1}}1} \tilde{e}^{\bar{2}}_{\ph{\bar{1}}2}\,, \\
 g_{22} &= \left( \tilde{e}^{\bar{2}}_{\ph{\bar{1}}2} \right)^2\,,
 \end{split}
\end{equation}
and
\begin{equation} \label{EEGU}
 \begin{split}
 g_{11} &= \left(\hat{e}^{\bar{1}}_{\ph{\bar{1}}1}\right)^2\,, \\
 g_{12} &= \hat{e}^{\bar{1}}_{\ph{\bar{1}}1} \hat{e}^{\bar{1}}_{\ph{\bar{1}}2}\,, \\
 g_{22} &= \left( \hat{e}^{\bar{2}}_{\ph{\bar{1}}2} \right)^2 + \left(\hat{e}^{\bar{1}}_{\ph{\bar{1}}2}\right)^2\,.
 \end{split}
\end{equation}
The metric is the ``physical'' field and thus the choice of vielbein is arbitrary and undetectable.\footnote{This is not true for fermions which couple to the vielbein.} Similarly, we expect that the 3-form and trivector fields of 11-dimensional supergravity should be treated on the same footing, and, in particular, that the generalised metric is the object that one should focus on as carrying the physical information, not the 3-form or trivector which are nothing but different parameterisations.

Now, we consider the Lorentzian case where we see that in geometry one may no longer be able to express the metric through an upper triangular or lower triangular vielbein. Since the two-dimensional Minkowski metric, $\eta_2$ is the internal flat metric, we decompose the metric into its vielbein via
\begin{equation}
 g = e^T \eta_2 e \,,
\end{equation}
so that equations \eqref{EEGL} and \eqref{EEGU} become
\begin{equation} \label{ELGL}
 \begin{split}
 g_{11} &= - \left(\tilde{e}^{\bar{1}}_{\ph{\bar{1}}1}\right)^2 + \left(\tilde{e}^{\bar{2}}_{\ph{\bar{1}}1}\right)^2\,, \\
 g_{12} &= \tilde{e}^{\bar{2}}_{\ph{\bar{1}}1} \tilde{e}^{\bar{2}}_{\ph{\bar{1}}2}\,, \\
 g_{22} &= \left( \tilde{e}^{\bar{2}}_{\ph{\bar{1}}2} \right)^2
 \end{split}
\end{equation}
and
\begin{equation} \label{ELGU}
 \begin{split}
 g_{11} &= - \left(\hat{e}^{\bar{1}}_{\ph{\bar{1}}1}\right)^2\,, \\
 g_{12} &= - \hat{e}^{\bar{1}}_{\ph{\bar{1}}1} \hat{e}^{\bar{1}}_{\ph{\bar{1}}2}\,, \\
 g_{22} &= \left( \hat{e}^{\bar{2}}_{\ph{\bar{1}}2} \right)^2 - \left(\hat{e}^{\bar{1}}_{\ph{\bar{1}}2}\right)^2\,.
 \end{split}
\end{equation}
Clearly by using a lower triangular vielbein $\tilde{e}$ we find that the component $g_{22} > 0$. Conversely, the metric can be described in terms of a lower triangular vielbein only if
\begin{equation}
 g_{22} > 0\,,
\end{equation}
while from equation \eqref{ELGU} we find that the metric can be described by an upper triangular vielbein $\hat{e}$ only if
\begin{equation}
 g_{11} < 0\,.
\end{equation}
Equivalently, an upper triangular vielbein $\hat{e}$ can be rotated into a lower triangular one, $\tilde{e} = H \hat{e}$, only if $g_{22} > 0$ and thus
\begin{equation}
 \left(\frac{\tilde{e}^{\bar{1}}_{\ph{\bar{1}}2}}{\tilde{e}^{\bar{2}}_{\ph{\bar{1}}2}}\right)^2 < 1\,.
\end{equation}
This is analogous to the statement for U-duality that the trivector can be gauged away only if $W^2 < 1$.

Thus we see that this problem of not being able to gauge away the trivector field arises generically in geometric constructions. It happens because we want to express our theory in terms of the metric and 3-form, $\left(g_{11}, C_3\right)$, but these are the ``wrong'' variables because they do not remain invariant under the local symmetry group $\tilde{H}_d$. The true physical field is the generalised metric $\gm$ which parameterises the appropriate coset
\begin{equation}
 \frac{E_{d} \times GL(\bar{d})}{\tilde{H}_d \times SO(\bar{d})}
\end{equation}
and remains invariant under the local symmetry group. One may in some cases express the generalised metric in terms of a metric and 3-form $\left(g_{11}, C_{3}\right)$ or a metric and trivector $\left(\og_{11}, \Omega_3\right)$ but not in general, just as the two-dimensional Lorentzian metric may in some cases be expressed as a lower triangular vielbein or an upper triangular one but in general neither.

We will see that there are four different scenarios that may arise. We classify the generalised metric of these scenarios as one of four types, depending on the signature of the two $3\times3$ matrices corresponding to the components $\gm_{ij}$ and $\gm^{ij,kl}$, as summarised in table \ref{TFrameConditions}.

\begin{table}
\centering
\begin{tabular}{|c|c|c||c|}
\hline
Type & $\gm_{ij}$ & $\gm^{ij,kl}$ & Minimal valid frames \\
\hline
I & $\left(-,+,+\right)$ & $\left(-,-,+\right)$ & $\left(g_{11}, C_3\right)$ or $\left(\og_{11}, \Omega_3\right)$ \\
II & $\left(+,-,-\right)$ or $0$ & $\left(-,-,+\right)$ & $\left(g_{11}, C_3\right)$ \\
III & $\left(-,+,+\right)$ & $\left(+,+,-\right)$ or $0$ & $\left(\og_{11}, \Omega_3\right)$ \\
IV & $\left(+,-,-\right)$ or $0$ & $\left(+,+,-\right)$ or $0$ & $\left(\hat{g}_{11}, C_3, \Omega_3\right)$ \\
\hline
\end{tabular}
\caption{The conditions for being able to use a certain field frame $\left(g_{11},C_3\right), \left(\og_{11},\Omega_3\right), \left(\hat{g}_{11},C_3,\Omega_3\right)$ in terms of the signature of the components of the generalised metric $\gm_{ij}$ and $\gm^{ij,kl}$. The minimal valid frames are those with the smallest number of bosonic fields that describe the physics. $\left(\hat{g}_{11},C_3,\Omega_3\right)$ is always a valid frame but we only include it when it is the only valid frame because it otherwise carries an unnecessary redundancy.}
\label{TFrameConditions}
\end{table}

\paragraph{Type I}
This corresponds to signatures
\begin{equation}
 \begin{split}
  \gm_{ij} & = \left(-,+,+\right) \,, \\
  \gm^{ij,kl} & = \left(-,-,+\right)\,.
 \end{split}
\end{equation}
We can use both the $\left(g_{11}, C_3\right)$ and $\left(\og_{11}, \Omega_3\right)$ frames.\footnote{We will choose one of the $\left(g_{11}, C_3\right)$ or $\left(\og_{11}, \Omega_3\right)$ frames whenever possible. One could, however, always use a frame including both a non-zero 3-form and trivector.}
The generalised metric and its generalised vielbein can correspondingly be written as either
\begin{equation}
 \begin{split}
  \gm &= |g_{11}|^{-1/2} \left( \begin{array}{ccc}
                               g \left(1 - V^2\right) & \frac{1}{\sqrt{2}} Cg^{-1} g^{-1} & 0 \\
			       \frac{1}{\sqrt{2}} g^{-1} g^{-1} C & g^{-1} g^{-1} & 0 \\
			       0 & 0 & g_8
                              \end{array} \right)\,, \\
  L_C &= |\tilde{e}_{11}|^{-1/2} \left( \begin{array}{ccc}
                                 \tilde{e} & 0 & 0 \\
				 \frac{1}{\sqrt{2}} e e C & e e & 0 \\
				 0 & 0 & \tilde{e}_8
                                \end{array} \right)\,,
 \end{split}
\end{equation}
or
\begin{equation}
 \begin{split}
  \gm &= |\og_{11}|^{-1/2} \left( \begin{array}{ccc}
                               \og & \frac{1}{\sqrt{2}} \og \Omega & 0 \\
			       \frac{1}{\sqrt{2}} \Omega \og & \og^{-1} \og^{-1} \left(1-W^2\right) & 0 \\
			       0 & 0 & \og_8
                              \end{array} \right)\,, \\
  L_\Omega &= |\tilde{\bar{e}}_{11}|^{-1/2} \left( \begin{array}{ccc}
                                 \tilde{\bar{e}} & \frac{1}{\sqrt{2}} \tilde{\bar{e}} \Omega & 0 \\
				 0 & \bar{e} \bar{e} & 0 \\
				 0 & 0 & \tilde{\bar{e}}_8
                                \end{array} \right)\,.
 \end{split}
\end{equation}
Because $\gm_{ij} = \left(-,+,+\right)$ we have $V^2 < 1$. Similarly, $\gm^{ij,kl} = \left(-,-,+\right)$ implies $W^2 < 1$. This means we can rotate the vielbeins from lower triangular to upper triangular by some $H \in SO(1,1)$ (see equations \eqref{ELTU1} - \eqref{3CtoOmegaD}) and this is why we can use both vielbeins.

\paragraph{Type II}
This corresponds to signatures
\begin{equation}
 \begin{split}
  \gm_{ij} &= \left(+,-,-\right) \textrm{ or } \gm_{ij} = 0\,, \\
  \gm^{ij,kl} & = \left(-,-,+\right)\,.
 \end{split}
\end{equation}
We can only use the $\left(g_{11}, C_3\right)$ frame. The generalised metric and its generalised vielbein are given by
\begin{equation}
 \begin{split}
  \gm &= |g_{11}|^{-1/2} \left( \begin{array}{ccc}
                               g \left(1 - V^2\right) & \frac{1}{\sqrt{2}} Cg^{-1} g^{-1} & 0 \\
			       \frac{1}{\sqrt{2}} g^{-1} g^{-1} C & g^{-1} g^{-1} & 0 \\
			       0 & 0 & g_8
                              \end{array} \right)\,, \\
  L_C &= |\tilde{e}_{11}|^{-1/2} \left( \begin{array}{ccc}
                                 \tilde{e} & 0 & 0 \\
				 \frac{1}{\sqrt{2}} e e C & e e & 0 \\
				 0 & 0 & \tilde{e}_8
                                \end{array} \right)\,.
 \end{split}
\end{equation}
In this case $\gm_{ij} = (+,-,-)$ or $\gm_{ij} = 0$, implying $V^2 \geq 1$, and thus we cannot rotate $L_C \nrightarrow L_\Omega = H L_C$. This confirms that we cannot rotate the 3-form away.

\paragraph{Type III}
This corresponds to
\begin{equation}
 \begin{split}
  \gm_{ij} & = \left(-,+,+\right)\,, \\
  \gm^{ij,kl} & = \left(+,+,-\right) \textrm{ or } \gm^{ij,kl} = 0\,.
 \end{split}
\end{equation}
We must use the $\left(\og_{11}, \Omega_3\right)$ frame. The generalised metric and its generalised vielbein are given by
\begin{equation}
 \begin{split}
  \gm &= |\og_{11}|^{-1/2} \left( \begin{array}{ccc}
                               \og & \frac{1}{\sqrt{2}} \og \Omega & 0 \\
			       \frac{1}{\sqrt{2}} \Omega \og & \og^{-1} \og^{-1} \left(1-W^2\right) & 0 \\
			       0 & 0 & \og_8
                              \end{array} \right)\,, \\
  L_\Omega &= |\tilde{\bar{e}}_{11}|^{-1/2} \left( \begin{array}{ccc}
                                 \tilde{\bar{e}} & \frac{1}{\sqrt{2}} \tilde{\bar{e}} \Omega & 0 \\
				 0 & \bar{e} \bar{e} & 0 \\
				 0 & 0 & \tilde{\bar{e}}_8
                                \end{array} \right)\,.
 \end{split}
\end{equation}
Now $\gm^{ij,kl} = \left(+,+,-\right)$ and so we find $W^2 > 1$ meaning we cannot rotate the upper triangular vielbein into a lower triangular one. Hence we cannot obtain the $\left(g_{11}, C_3\right)$ frame.

\paragraph{Type IV}
This arises when
\begin{equation}
 \begin{split}
  \gm_{ij} & = \left(+,-,-\right) \textrm{ or } \gm_{ij} = 0 \,, \\
  \gm^{ij,kl} & = \left(+,+,-\right) \textrm{ or } \gm^{ij,kl} = 0 \,, 
 \end{split}
\end{equation}
and we have to use both a 3-form and a trivector. There are in fact two frames: $\left(\hg_{11}, \hat{C}_3, \hat{\Omega}_3\right)$ and $\left(\cg_{11}, \check{C}_3, \check{\Omega}_3\right)$. Corresponding to each of these frames we can parameterise the generalised metric and its generalised vielbein in one of two forms. The hatted frame gives
\begin{equation}
 \begin{split}
  \gm &= |\hg_{11}|^{-1/2} \left( \begin{array}{ccc}
               \hg \left[ \left(1 - \hat{W}\hat{V}\right)^2 - \hat{V}^2\right] & \frac{1}{\sqrt{2}} \left[\hg \hat{\Omega} \left(1-\hat{W} \hat{V} \right) + \hat{C} \hg^{-1} \hg^{-1} \right] & 0 \\
		\frac{1}{\sqrt{2}} \left[ \hat{\Omega} \hg \left(1 - \hat{W} \hat{V} \right) + \hg^{-1} \hg^{-1} \hat{C} \right] & \hg^{-1} \hg^{-1} \left(1 - \hat{W}^2\right) & 0 \\
		0 & 0 & \hg_8 \end{array}  \right)\,, \\
  \hat{L} &= |\tilde{\hat{e}}|^{-1/2}
           \left( \begin{array}{ccc}
           \tilde{\hat{e}} \left(1 - \hat{W} \hat{V}\right) & \frac{1}{\sqrt{2}} \tilde{\hat{e}} \hat{\Omega} & 0 \\
           \frac{1}{\sqrt{2}} \hat{e} \hat{e} \hat{C} & \hat{e} \hat{e} & 0 \\
	   0 & 0 & \hat{\tilde{e}}_8
           \end{array} \right)\,.
 \end{split}
\end{equation}
The generalised vielbein is fixed by the requirement that it is a group element of local $E_{3} \times GL(8)$ so it must be formed by
\begin{equation}
 \hat{L} = U_\Omega(X) U_C(X) U_\alpha(X) U_{SL(3)}(X) U_{GL(8)}(X)\,,
\end{equation}
where each factor is an element of $E_{3} \times GL(8)$ as given in \eqref{ESL(d)}, \eqref{ESL(2)1} - \eqref{ESL(2)3} and is a function of the generalised coordinates. We can interpret each factor as turning on a specific field, in particular $U_\alpha U_{SL(3)}$ turns on the gravitational field in the dualisable direction while $U_{GL(8)}$ turns it on in the transverse space, while $U_\Omega$ and $U_C$ turn on the trivector and 3-form, respectively. However, because these elements are constructed from the duality algebra, the trivector and 3-form obtained this way have tangent space indices and thus they must always be on the left of the gravitational field factors $U_\alpha U_{SL(3)} U_{GL(8)}$. These last three factors commute and thus their order does not matter. On the other hand, we could change the order of $U_\Omega U_C$. This gives rise to the generalised vielbein
\begin{equation}
 \check{L} = U_C(X) U_\Omega(X) U_\alpha(X) U_{SL(3)}(X) U_{GL(8)}(X)\,.
\end{equation}
This is the checked frame's vielbein
\begin{equation}
 \check{L} = |\tilde{\check{e}}|^{-1/2}
           \left( \begin{array}{ccc}
           \tilde{\check{e}} & \frac{1}{\sqrt{2}} \tilde{\check{e}} \check{\Omega} & 0 \\
           \frac{1}{\sqrt{2}} \check{e} \check{e} \check{C} & \check{e} \check{e}  \left(1 - \check{W} \check{V}\right) & 0 \\
	   0 & 0 & \check{\tilde{e}}_8
           \end{array} \right)\,,
\end{equation}
with generalised metric
\begin{equation}
 \gm = |\cg_{11}|^{-1/2} \left( \begin{array}{ccc}
                                 \cg \left(1 - \check{V}\right)^2 & \frac{1}{\sqrt{2}} \left[ \check{C} \cg^{-1} \cg^{-1} \left( 1 - \check{V} \check{W} \right) + \cg \check{\Omega} \right] & 0 \\
				 \frac{1}{\sqrt{2}} \left[ \cg^{-1} \cg^{-1} \check{C} \left( 1 - \check{V} \check{W} \right) + \check{\Omega} \cg \right] & \cg^{-1} \cg^{-1} \left[ \left(1 - \check{V} \check{W} \right)^2 - \check{W}^2 \right] & 0 \\
				 0 & 0 & \cg_8
                                \end{array} \right)\,.
\end{equation}

However, it is easy to check that these two frames are related by the field redefinition
\begin{equation}
 \begin{split}
  \cg & = \hg \left|1 - \hat{V} \hat{W}\right|^{4/3}\,, \\
  \cg_8 &= \hg_8 \left| 1 - \hat{V} \hat{W} \right|^{-2/3}\,, \\
  \check{\Omega}_3 &= \hat{\Omega}_3 \left(1-\hat{V}\hat{W}\right)^{-1}\,, \\
  \check{C}_3 &= \hat{C}_3 \left(1-\hat{V}\hat{W}\right)\,, \\
  \check{V} &= \hat{V} \left(1 - \hat{V} \hat{W} \right)\,, \\
  \check{W} &= \hat{W} \left(1 - \hat{V} \hat{W} \right)^{-1}\,, \\
  \tilde{\check{e}} &= \tilde{\hat{e}} \left(1-\hat{V}\hat{W}\right)^{2/3}\,, \\
  \tilde{\check{e}}_8 &= \tilde{\hat{e}}_8 \left(1-\hat{V}\hat{W}\right)^{-1/3}\,. \\
 \end{split} \label{EHatToCheck}
\end{equation}

It is important to note that the metric, 3-form and trivector appearing in this generalised metric are not unique. For a start, we can use the field redefinitions \eqref{EHatToCheck} to obtain an equally valid set of fields. Also, the structure of $\hat{L}$ is preserved by all internal rotations $H$ of the form
\begin{equation}
 H = \left( \begin{array}{ccc}
            \cosh \theta \delta^{\bar{i}}_{\ph{\bar{i}}\bar{k}} & \frac{1}{\sqrt{2}} \sinh \theta \epsilon^{\bar{i}\bar{k}\bar{l}} & 0 \\
            -\frac{1}{\sqrt{2}} \sinh \theta \epsilon_{\bar{i}\bar{j} \bar{k}} & \cosh \theta \delta_{\bar{i}\bar{j}}^{\ph{\bar{i}\bar{j}}\bar{k}\bar{l}} & 0 \\
	    0 & 0 & 1
            \end{array} \right)\,.
\end{equation}
Because the metric is of type IV there are no values for $\theta$ which turn the vielbein into a lower triangular $L_C$ or upper triangular $L_\Omega$ one.



\section{Timelike dualities and change of signature} \label{STU}
We will now review why it seems that M-theory changes signature under the action of timelike dualities \cite{Hull:1998vg,Hull:1998ym,Hull:1998br}. Conventionally, dualities arise when considering compactifications. We start by compactifying 11-dimensional supergravity on a $S^1$ of radius $R_{1}$ and take the limit $R_{1} \rightarrow 0$ to obtain the type IIA 10-dimensional supergravity. A Kaluza-Klein Ansatz for the compactification shows that the dilaton is related to this radius $e^{\phi} = R_1^{3/2}$ so that we are considering the weak-coupling limit \cite{Witten:1995ex}. A further compactification on a circle of radius $R_2$ gives the T-dual IIB supergravity compactified on a dual circle of radius $1/R_2$. Thus 11-dimensional supergravity compactified on $T^2$ in the limit of $R_1, R_2 \rightarrow 0$ is dual to a 10-dimensional supergravity. From this analysis we see that for every two-cycle we compactify on, we get a dimension opening up in the dual theory when the two-cycle shrinks to zero size. Thus a compactification of 11-dimensional supergravity on $T^3$ is dual to a $11 - 3 + 1\times3 = 11$ dimensional theory when the $T^3$ has vanishing size because $T^3$ has three two-cycles.

Let us now see what happens when we compactify on Lorentzian torii. We denote by $T^{(n,p)}$ the torus with $n$ spacelike and $p$ timelike directions. Now we consider compactifying the 11-dimensional supergravity on a $T^{(1,1)}$. We use the spacelike circle to obtain the IIA theory (in the limit of vanishing radius $R_1$) whereas the compactification on the timelike circle of radius $R_2$ relates the theory to a 10-dimensional theory compactified on a timelike circle of dual radius $1/R_2$. In the limit of vanishing size we see that while one spatial and one timelike direction disappear in the original solution a timelike one opens up in the dual spacetime. Thus for every Lorentzian two-cycle that we compactify on we open up a dual timelike direction whereas for each Euclidean two-cycle we open up a dual spacelike direction. We summarise

\begin{equation*}
 \begin{array}{cc}
  \textrm{Compactification } & \textrm{ Dual spacetime} \\
  \hline
  \textrm{Each shrinking Euclidean 2-cycle } &\rightarrow \textrm{ New spacelike direction opens up} \\
  \textrm{Each shrinking Lorentzian 2-cycle } &\rightarrow \textrm{ New timelike direction opens up.}
 \end{array}
\end{equation*}

Now when we consider a compactification of 11-dimensional supergravity on $T^{(2,1)}$ we go from a $(1,10)$ theory to a $(1, 10) - (1, 2) + 2\times (1,0) + (0,1) = (2, 9)$ theory, denoted by M$^*$. This is because the $T^{(2,1)}$ has two Lorentzian two-cycles and one Euclidean two-cycle.

Using the notation of generalised geometry we naively get the same results. We see that if we perform a Buscher duality, equation \eqref{EBuscherGenerated}, along three directions $t, x^1, x^2$ we will exchange the spacetime coordinates with their duals because
\begin{equation}
 X \rightarrow \left(U_B\right)^{-1} X\,,
\end{equation}
where $X$ are the generalised coordinates and
\begin{equation}
 U_B = \left( \begin{array}{ccc}
               0 & \frac{1}{\sqrt{2}} D & 0 \\
	       \frac{1}{\sqrt{2}} K & 0 & 0 \\
	       0 & 0 & 1
              \end{array} \right)\,,
\end{equation}
where $D^{t12} = A$ and $K_{t12} = -\frac{1}{A}$. Explicitly we have for $A = 1$ (we will set $A = 1$ throughout this section unless specified otherwise)
\begin{equation}
 \begin{split}
  t &\leftrightarrow y_{12} \,, \\
  x^1 & \leftrightarrow - y_{t2} \,, \\
  x^2 & \leftrightarrow y_{t1}\,.
 \end{split}
\end{equation}
Equation \eqref{ETA3} revealed that the dual coordinates $y_{t\mu}$ for $\mu = 1, 2$ are timelike while only $y_{12}$ is spacelike. Thus, we seem to obtain the same result as in \cite{Hull:1998ym} obtaining a dual theory of signature $(9,2)$.

However, let us study this more carefully using the generalised metric. It can be contracted with the generalised coordinates to give a U-duality invariant generalised line element
\begin{equation}
 dS^2 = \gm_{MN} dX^M dX^N\,.
\end{equation}
For vanishing 3-form this simplifies to
\begin{equation}
 dS^2 = g_{ij} dx^i dx^j + \frac{1}{2} g^{ik}g^{jl} dy_{ij} dy_{kl} + g_{AB} dx^A dx^B \,,
\end{equation}
and by studying the line element restricted along the spacetime coordinates, $ds^2 = g_{ab} dx^a dx^b$, we can obtain the metric. After applying $U_B$ the spacetime coordinates are now made up of two ``timelike'', $y_{t1}, y_{t2}$, and nine ``spacelike'' coordinates, $y_{12}, x^3, \ldots x^{10}$, and we would expect the metric to have changed signature. Implicitly we are assuming that the 3-form vanishes. In the Euclidean case that would be true. Under a Buscher duality along spacelike directions the fields transform as \eqref{3BuscherDuality}
\begin{equation}
 \begin{split}
  g'_{ij} &= g_{ij} \left(\left( C_{123}^2 + |g_3| \right) \right)^{-2/3}\,, \\
  g'_{AB} &= g_{AB} \left(\left( C_{123}^2 + |g_3| \right) \right)^{1/3}\,, \\
  C'_{123} &= - \frac{C_{123}}{\left( C_{123}^2 + |g_3| \right)}\,,
 \end{split}
\end{equation}
and it would be true that there is no dual 3-form if we started with a vanishing 3-form.

However, the generalised metric in the Lorentzian case transforms as
\begin{align}
 \gm' &= |g_{11}|^{-1/2} \left(U_B\right)^T \left( \begin{array}{ccc}
                              g & 0 & 0 \\
                              0 & g^{-1} g^{-1} & 0 \\
                              0 & 0 & g_8
                              \end{array} \right) U_B \\
 &= |g_{11}|^{-1/2} \left( \begin{array}{ccc}
                            0 & \frac{1}{\sqrt{2}} D & 0 \\
                            \frac{1}{\sqrt{2}} K & 0 & 0 \\
                            0 & 0 & 1
                            \end{array} \right)
                            \left( \begin{array}{ccc}
                              g & 0 & 0 \\
                              0 & g^{-1} g^{-1} & 0 \\
                              0 & 0 & g_8
                              \end{array} \right)
                            \left( \begin{array}{ccc}
                            0 & \frac{1}{\sqrt{2}} K & 0 \\
                            \frac{1}{\sqrt{2}} D & 0 & 0 \\
                            0 & 0 & 1
                            \end{array} \right) \\
 &= |g_{11}|^{-1/2} \left( \begin{array}{ccc}
                                      -\frac{1}{|g|} g & 0 & 0 \\
                                      0 & -|g| g^{-1} g^{-1} & 0 \\
                                      0 & 0 & g_8
                                      \end{array} \right)\,.
\end{align}
We see that the naive interpretation, that the metric along dualisable directions $g_{ij}$ has reversed signature $g_{ij} \rightarrow - g_{ij}$, so that it now has two timelike and one spacelike direction, is incorrect. Because $\gm'^{ij,kl}$ has signature $\left(+,+,-\right)$ the generalised metric is now of type IV and so we need to also include a trivector field. As we will explain in the next section, we find the dual fields
\begin{equation}
 \begin{split}
  \cg &= g \left(\sinh \theta\right)^{4/3} |g|^{2/3}\,, \\
  \cg_{8} &= g_8 \left(\sinh\theta\right)^{-2/3} |g|^{-1/3}\,, \\
  \check{C}_{t12} &= - \frac{\cosh\theta\sinh\theta}{\sqrt{|g|}}\,, \\
  \check{\Omega}^{t12} &= \sqrt{|g|} \coth \theta\,,
 \end{split}
\end{equation}
where $A \sinh \theta \geq 0$ is required, i.e. $\theta$ has to be chosen to be the same sign as $A$. We see that there is no change in signature. However, there is a trivector field and a family of dual solutions, linked by local $SO(1,1)$ rotations. We emphasise that it is the existence of the trivector field that saves us from a change of signature.

\subsection{The spacetime signature}
We can go further and prove the following theorem.
\begin{thm}
If the generalised metric parameterises the coset
\begin{equation}
 \frac{SL(2) \times SL(3) \times GL(8)}{SO(1,1) \times SO(2,1) \times SO(8)}\,,
\end{equation}
then the spacetime metric must be of signature $\left(-,+, \ldots, +\right)$.
\end{thm}
\paragraph{Proof}
The generalised metric is symmetric and can thus be written in terms of a generalised vielbein
\begin{equation}
 \gm = L^T \mathcal{M} L\,.
\end{equation}
As we have shown in section \ref{STGenGeom}, the internal metric is fixed by the local symmetry group $SO(1,1) \times SO(2,1) \times SO(8)$ to be
\begin{equation}
 \mathcal{M} = diag \left(-, +, +, -, -, +\right) \otimes 1_8\,,
\end{equation}
where the pseudo-Riemannian part is made from the components
\begin{align}
 \mathcal{M}_{ij} &= \eta_{ij}\,, \\
 \mathcal{M}^{ij,kl} &= \eta^{i[k} \eta^{l]j} \,,
\end{align}
and the components for the transverse space are $\mathcal{M}_{AB} = \delta_{AB}$. Here $\eta$ is the three-dimensional Minkowski metric.

The generalised vielbein has to be a group element and thus must be of the form
\begin{equation}
 L = U_{\Omega}(X) U_C(X) U_\alpha(X) U_{SL(3)}(X) U_{GL(8)}(X)\,,
\end{equation}
where each factor is a function of the generalised coordinates $X^M = \left(x^i, y_{ij}, x^A\right)$. The indices are as usual $i, j = 1, 2, 3$ and $A = 4, \ldots 11$. We have shown in section \ref{STIntSym} that this form is generic because the alternative,
\begin{equation}
 L = U_C(X) U_\Omega(X) U_\alpha(X) U_{SL(3)}(X) U_{GL(8)}(X)\,,
\end{equation}
can be obtained by the field redefinitions given in equations \eqref{EHatToCheck}. Thus, we can without loss of generality write the vielbein as
\begin{equation}
 L = |\tilde{e}|^{-1/2}
           \left( \begin{array}{ccc}
           \tilde{e} \left(1 - W V\right) & \frac{1}{\sqrt{2}} \tilde{e} \Omega & 0 \\
           \frac{1}{\sqrt{2}} e e C & e e & 0 \\
     	   0 & 0 & \tilde{e}_8
           \end{array} \right)\,.
\end{equation}
The generalised metric is given by
\begin{equation}
  \gm = |g_{11}|^{-1/2} \left( \begin{array}{ccc}
        g \left[ \left(1 - WV\right)^2 - V^2\right] & \frac{1}{\sqrt{2}} \left[g \Omega \left(1-W V \right) + C g^{-1} g^{-1} \right] & 0 \\
		\frac{1}{\sqrt{2}} \left[ \Omega g \left(1 - W V \right) + g^{-1} g^{-1} C \right] & g^{-1} g^{-1} \left(1 - W^2\right) & 0 \\
		0 & 0 & g_8 \end{array}  \right)\,.
\end{equation}
For now we take $g_{11} = g_ \otimes g_8, C_3$ and $\Omega_3$ to be some symmetric rank-two field, a 3-form and a trivector, respectively, each of unknown physical significance. $g$ is given by
\begin{equation}
 g = \tilde{e}^T \eta \tilde{e}\,,
\end{equation}
and
\begin{equation}
 g_8 = \tilde{e}_8^{\ph{8}T} \tilde{e}_8\,.
\end{equation}
When $\Omega_3 = 0$, the fields $g_{11}$ and $C_3$ are the spacetime metric and
3-form, respectively. They have to be because the low-energy effective action
\eqref{ELEEA} must reduce to the Einstein-Hilbert action when $\partial_{y} =
0$. Also, the generalised metric can be found by considering the action of
dualities on the worldvolume of the supermembrane \cite{Duff:1990hn} and by
comparison we see that $g_{11}$ and $C_3$ are the usual bosonic fields of
11-dimensional supergravity. By continuity $g_{11}$ must be the spacetime
metric when the 3-form and trivector are non-vanishing. Thus the spacetime
internal metric $\eta \otimes 1_8$ determines the spacetime signature to be
$\left(-, +, +, +, \ldots +\right)$.

By a similar argument one can prove the relevant theorem for the Lorentzian modular group of the four-dimensional duality group $\frac{E_4 \times GL(7)}{\tilde{H}_4 \times SO(7)}$ as given in table \ref{TDualityGroups}.
\begin{thm}
If the generalised metric parameterises the coset
\begin{equation}
 \frac{SL(5) \times GL(7)}{SO(3,2) \times SO(7)}\,,
\end{equation}
then the spacetime metric must be of signature $\left(-,+, \ldots, +\right)$\,.
\end{thm}

\section{The transformation rules} \label{STDuality}
We can now repeat the analysis in \cite{Malek:2012pw} including time amongst the dualisable coordinates in order to find the bosonic fields after the action of a duality. We will start with a type I or type II generalised metric so that we can use the $\left(g_{11}, C_3\right)$ frame. Including time means that the dual generalised metric may have changed type and thus the dual fields may not be expressible in the $\left(g_{11},C_3\right)$ frame.

The non-trivial dualities are generated by the $SL(2)$ subgroup
\begin{equation}
 \left\{U_C, U_\Omega, U_\alpha\right\}\,.
\end{equation}
In the $\left(g_{11}, C_3\right)$ frame the $U_C$ shifts the 3-form and $U_\alpha$ always scales the coordinates. Thus, these two transformations are clearly gauge transformations. However, $U_\Omega$ transforms the bosonic fields in a non-trivial manner. Another non-trivial transformation is generated by the Buscher duality $U_B$
\begin{equation}
 U_B = U_C U_\Omega U_C\,,
\end{equation}
where $\Omega^{t12} = A$ and $C_{t12} = -\frac{1}{A}$. We consider their action on the bosonic fields in turn.

\subsection{$\Omega$-shifts}
We start with the generalised metric
\begin{equation}
 \gm = |g_{11}|^{-1/2}
        \left( \begin{array}{cc}
        g \left(1 - V^2\right) & \frac{1}{\sqrt{2}} C g^{-1} g^{-1} \\
        \frac{1}{\sqrt{2}} g^{-1} g^{-1} C & g^{-1} g^{-1}
        \end{array} \right)\,. \label{GMC}
\end{equation}
Applying a $U_\Omega$ transformation we find
\begin{equation}
 \gm' = |g_{11}|^{-1/2}
        \left( \begin{array}{cr}
        g \left(1 - V^2\right) & \frac{1}{\sqrt{2}} C g^{-1} g^{-1} \left( 1 + A \sqrt{|g|} \left(\frac{1}{V}-V\right) \right) \\
        \frac{1}{\sqrt{2}} g^{-1} g^{-1} C \left( 1 + A \sqrt{|g|} \left(\frac{1}{V}-V\right) \right) & g^{-1} g^{-1} \left( \left(1 - A \sqrt{g} V \right)^2 - A^2 |g| \right)
        \end{array} \right)\,,
\end{equation}
where $\Omega^{t12} = A$, so that $\Omega^{ijk} = \epsilon^{ijk} \sqrt{|g|} A$. Equivalently, we can write
\begin{equation}
 \gm' = |g_{11}|^{-1/2}
        \left( \begin{array}{cc}
        g \left(1 - V^2\right) & \frac{1}{\sqrt{2}} C g^{-1} g^{-1} \left( 1 + A C_{t12} \left(\frac{1}{V^2}-1\right) \right) \\
        \frac{1}{\sqrt{2}} g^{-1} g^{-1} C \left( 1 + A C_{t12} \left(\frac{1}{V^2}-1\right) \right) & g^{-1} g^{-1} \left( \left(1 + A C_{t12} \right)^2 - A^2 |g| \right)
        \end{array} \right)\,.
\end{equation}

We see that $\gm'^{ij,kl}$ may reverse signature. If
\begin{equation}
 \left(1 + A C_{t12} \right)^2 - A^2 |g| > 0\,, \label{EOGOA}
\end{equation}
we can gauge away the trivector field and find that in the $\left(g_{11}, C_3\right)$ frame, the transformed fields are
\begin{equation}
 \begin{split}
 g'_{ij} &= g_{ij} \left( \left(1 + A C_{t12}\right)^2 - A^2 |g| \right)^{-2/3}\,, \\
 g'_{AB} &= g_{AB} \left( \left(1 + A C_{t12}\right)^2 - A^2 |g| \right)^{1/3}\,, \\
 C'_{t12} &= \frac{C_{t12} \left( 1 + A C_{t12}\right) - A |g|}{\left(1 + A C_{t12}\right)^2 - A^2 |g|}\,. \label{ETOmega}
 \end{split}
\end{equation}
However, if
\begin{equation}
 \left(1 + A C_{t12} \right)^2 - A^2 |g| \leq 0\,, \label{EONGOA}
\end{equation}
$\gm'^{ij,kl}$ has reversed signature and the trivector field cannot be gauged away since the dual generalised metric is of type III or IV. If $V^2 < 1$ the dual generalised metric is of type III and we can gauge away the 3-form. To find the dual fields, we first gauge away the initial 3-form away so the initial fields are in the $\left(g_{11}, \Omega_3\right)$ frame given in equations \eqref{3CtoOmegaD}, and then add the trivector $\Omega^{t12} = A$.
\begin{equation}
 \begin{split}
 \og_{ij} &= g_{ij} \left(1 - V^2\right)^{2/3}\,, \\
 \Omega^{ijk} &= \frac{\epsilon^{ijk}V}{1-V^2} + A \sqrt{|g|} \epsilon^{ijk}= \frac{g^{im} g^{jn} g^{ko} C_{mno}}{1-V^2} + A \sqrt{|g|} \epsilon^{ijk}\,, \\
 \og_{AB} &= g_{AB} \left(1 - V^2\right)^{-1/3} \, .
 \end{split}
\end{equation}
On the other hand, if $V^2 \geq 1$ the dual generalised metric is of type IV and we have to use the 3-form and trivector. In the $\left(\cg_{11}, \check{C}_3, \check{\Omega}_3\right)$ frame, the trivector can just be added to the metric and 3-form.
\begin{equation}
 \begin{split}
  \cg'_{ab} &= g_{ab}\,, \\
  \check{C}'_{t12} &= C_{t12}\,, \\
  \check{\Omega}'^{t12} &= A\,.
 \end{split}
\end{equation}
In this frame fields linked by a $SO(1,1)$ rotation are equally valid. We thus find a family of dual solutions given by
\begin{equation} \label{ETOmegaCheck}
 \begin{split}
  \cg'_{ij} &= g_{ij} \left(\cosh\theta + \frac{C_{t12}}{\sqrt{|g|}} \sinh\theta\right)^{4/3}\,, \\
  \cg'_{AB} &= g_{AB} \left(\cosh\theta + \frac{C_{t12}}{\sqrt{|g|}} \sinh\theta\right)^{-2/3}\,, \\
  \check{C}'_{t12} &= \sqrt{|g|} \left(\cosh\theta + \frac{C_{t12}}{\sqrt{|g|}} \sinh\theta\right) \left(\frac{C_{t12}}{\sqrt{|g|}} \cosh\theta + \sinh\theta\right)\,, \\
  \check{\Omega}'^{t12} &= \frac{A \sqrt{|g|} \cosh\theta + \sinh\theta \left(1 + A C_{t12} \right)}{\sqrt{|g|} \cosh\theta + C_{t12} \sinh\theta}\,,
 \end{split}
\end{equation}
which is valid for all $\theta$ satisfying $\cosh\theta > V\sinh\theta$. We highlight that the hyperbolic angle can be chosen locally, $\theta = \theta(X)$. One can also use the hatted frame by the field redefinition \eqref{EHatToCheck}.
\begin{equation}
 \begin{split} \label{ETOmegaHat}
  \hg'_{ij} & = g_{ij} \left[A \sqrt{|g|} \sinh\theta + \cosh\theta \left(1 + A C_{t12}\right)\right]^{-4/3}\,, \\
  \hg'_{AB} &= g_{AB} \left[A \sqrt{|g|} \sinh\theta + \cosh\theta \left(1 + A C_{t12}\right)\right]^{2/3}\,, \\
  \hat{\Omega}'^{t12} &= \frac{1}{\sqrt{|g|}} \left[A \sqrt{|g|} \cosh\theta + \sinh\theta \left( 1 + A C_{t12} \right) \right] \left[A \sqrt{|g|} \sinh\theta + \cosh\theta \left(1 + A C_{t12}\right)\right]\,, \\
  \hat{C}'_{t12} &= \frac{C_{t12} \cosh\theta + \sqrt{|g|} \sinh\theta}{A \sqrt{|g|} \sinh\theta + \cosh\theta \left(1 + A C_{t12}\right)}\,.
 \end{split}
\end{equation}

We can check that if $\left(1 + A C_{t12}\right)^2 > A^2 |g|$, we can rotate away the trivector field and obtain the fields in the $\left(g_{11}, C_3\right)$ frame as in equation \eqref{ETOmega}. We need to choose
\begin{equation}
 \cosh \theta = - \sinh\theta \frac{1 + AC_{t12}}{A\sqrt{|g|}}\,,
\end{equation}
which then implies
\begin{equation}
 \sinh\theta = \textrm{sign}\left(\frac{1+AC_{t12}}{-A\sqrt{|g|}}\right) \sqrt{\frac{A^2|g|}{\left(1+AC_{t12}\right)^2 - A^2|g|}}\,.
\end{equation}
For this choice it is easy to check that indeed the fields in both the hatted and checked frames reduce as required to
\begin{equation}
 \begin{split}
 g'_{ij} &= g_{ij} \left( \left(1 + A C_{t12}\right)^2 - A^2 |g| \right)^{-2/3}\,, \\
 g'_{AB} &= g_{AB} \left( \left(1 + A C_{t12}\right)^2 - A^2 |g| \right)^{1/3}\,, \\
 C'_{t12} &= \frac{C_{t12} \left( 1 + A C_{t12}\right) - A |g|}{\left(1 + A C_{t12}\right)^2 - A^2 |g|}\,.
 \end{split}
\end{equation}

\subsection{Buscher duality}
The other non-trivial duality is the Buscher duality
\begin{equation}
 U_B = U_C U_\Omega U_C\,,
\end{equation}
where $\Omega^{t12} = A$ and $C_{t12} = -\frac{1}{A}$. In \cite{Malek:2012pw} we calculated the effect of this duality by changing frames from $\left(g_{11}, C_3\right)$ to $\left(\og_{11}, \Omega_3\right)$ and back so that at each step the transformation is just a simple gauge shift
\begin{equation}
 \begin{split}
  C_{t12} & \rightarrow C_{t12} - \frac{1}{A}\,, \\
  \Omega^{t12} & \rightarrow \Omega^{t12} + A\,,
 \end{split}
\end{equation}
etc. However, here we need a different approach as we cannot always change frames from $\left(g_{11}, C_3\right)$ to $\left(\og_{11}, \Omega_3\right)$ or vice versa. We must study the transformation of the generalised metric directly. We first write the Buscher transformation as
\begin{equation}
 U_B = \left( \begin{array}{cc}
               0 & \frac{1}{\sqrt{2}} D \\
	       \frac{1}{\sqrt{2}} K & 0
              \end{array} \right)\,,
\end{equation}
where $D^{t12} = A$ and $K_{t12} = -\frac{1}{A}$. We start again with a type I or II generalised metric expressible in the $\left(g_{11}, C_3\right)$ frames. Under this transformation the generalised metric becomes
\begin{align}
 \gm' &= U_B^{\ph{B}T} \gm U_B \nonumber \\
      &= \left(-1\right)|g_{11}|^{-1/2}
	\left( \begin{array}{cc}
        A^{-2}|g|^{-1} g & \frac{1}{\sqrt{2}} C g^{-1} g^{-1} \\
	\frac{1}{\sqrt{2}} g^{-1} g^{-1} C & A^2 |g| \left(1-V^2\right) g^{-1} g^{-1}
        \end{array} \right)\,. \label{EGMBuscher}
\end{align}
We note that because of the $-1$ pre-multiplying the generalised metric, the component
\begin{equation}
 \gm'_{ij} = - |g_{11}|^{-1/2} A^{-2} |g|^{-1} g
\end{equation}
always has the reversed signature, $(+, -, -)$. We therefore always have to use a 3-form, $C'_{ijk}$. If $V^2 > 1$ then the dual generalised metric is of type II and we can use the frame $\left(g_{11}, C_3\right)$ with the dual fields given by the ``timelike Buscher rules'' \cite{Buscher:1987sk,Buscher:1987qj}
\begin{equation}
 \begin{split}
  g'_{ij} &= g_{ij} A^{-4/3} \left(C_{t12}^2 - |g| \right)^{-2/3}\,, \\
  g'_{AB} &= g_{AB} A^{2/3} \left( C_{t12}^2 - |g| \right)^{1/3}\,, \\
  C'_{t12} & = - \frac{C_{t12}}{A^2 \left(C_{t12}^2 - |g| \right)}\,. \label{ETBuscher}
 \end{split}
\end{equation}
If, on the other hand, $V^2 \leq 1$, the generalised metric is of type IV and we must include a non-zero trivector. The generalised vielbein becomes
\begin{equation}
 L' = |\tilde{e}_{11}|^{-1/2} \left( \begin{array}{cc}
              0 & \frac{1}{\sqrt{2}} \tilde{e} D \\
	      \frac{1}{\sqrt{2}} ee K & - ee V A \sqrt{|g|}
             \end{array} \right) \,,
\end{equation}
which can be rotated into the checked frame
\begin{equation}
 \begin{split}
  \check{L} &= |\tilde{\check{e}}|^{-1/2}
           \left( \begin{array}{cc}
           \tilde{\check{e}} & \frac{1}{\sqrt{2}} \tilde{\check{e}} \check{\Omega} \\
           \frac{1}{\sqrt{2}} \check{e} \check{e} \check{C} & \check{e} \check{e}  \left(1 - \check{W} \check{V}\right)
           \end{array} \right) \\
  &= H L'\,,
 \end{split}
\end{equation}
where $H \in SO(1,1)$ is given by equation \eqref{EHInt}
\begin{equation}
 H = \left( \begin{array}{cc}
            \cosh \theta \delta^{\bar{i}}_{\ph{\bar{i}}\bar{k}} & - \frac{1}{\sqrt{2}} \sinh \theta \epsilon^{\bar{i}\bar{k}\bar{l}} \\
            \frac{1}{\sqrt{2}} \sinh \theta \epsilon_{\bar{i}\bar{j}\bar{k}} & \cosh \theta \delta_{\bar{i}\bar{j}}^{\bar{k}\bar{l}}
            \end{array} \right)\,.
\end{equation}
We find the dual solutions belonging to a family of solutions linked by internal $SO(1,1)$ rotations (recall that the parameter $\theta(X)$ can be chosen locally)
\begin{equation} \label{ETBuscherCheck}
 \begin{split}
  \check{g}'_{ij} &= g_{ij} \left(\frac{\sinh\theta}{A\sqrt{|g|}} \right)^{4/3}\,, \\
  \check{g}'_{AB} &= g_{AB} \left(\frac{\sinh\theta}{A\sqrt{|g|}} \right)^{-2/3}\,, \\
  \check{C}'_{t12} &= - \frac{\sinh2\theta}{2 A^2 \sqrt{|g|}}\,, \\
  \check{\Omega}'^{t12} &= A^2 \sqrt{|g|} \left(\coth \theta - \frac{C_{t12}}{\sqrt{|g|}} \right)\,,
 \end{split}
\end{equation}
where $A \sinh\theta > 0$. We can also write the family of dual solutions in the hatted frame
\begin{equation} \label{ETBuscherHat}
 \begin{split}
  \hg'_{ij} &= g_{ij} \left(A\sqrt{|g|}\right)^{-4/3} \left| \sinh\theta - \frac{C_{t12}}{\sqrt{|g|}} \cosh\theta \right|^{-4/3}\,, \\
  \hg'_{AB} &= g_{AB} \left(A\sqrt{|g|}\right)^{2/3} \left| \sinh\theta - \frac{C_{t12}}{\sqrt{|g|}} \cosh\theta \right|^{2/3}\,, \\
  \hat{C}'_{t12} &= \left[A^2 \sqrt{|g|} \left(\tanh\theta - \frac{C_{t12}}{\sqrt{|g|}}\right)\right]^{-1}\,, \\
  \hat{\Omega}'^{t12} &= - A^2 \sqrt{|g|} \left(\cosh\theta - \frac{C_{t12}}{\sqrt{|g|}} \sinh\theta\right)\left(\sinh\theta - \frac{C_{t12}}{\sqrt{|g|}} \cosh\theta\right)\,.
 \end{split}
\end{equation}

Just as for the $\Omega$-shift, it is worth checking that if $V^2 > 1$ we can rotate the trivector field away. This would correspond to the choice
\begin{equation}
 V \sinh\theta = \cosh\theta\,,
\end{equation}
so that
\begin{equation}
 \sinh\theta = \textrm{sign}(V) \frac{1}{\sqrt{V^2-1}}\,,
\end{equation}
and the fields in both frames collapse to $\left(g_{11}, C_3\right)$ as expected
\begin{equation}
 \begin{split}
  g'_{ij} &= g_{ij} A^{-4/3} \left(C_{t12}^2 - |g|\right)^{-2/3}\,, \\
  g'_{AB} &= g_{AB} A^{2/3} \left(C_{t12}^2 - |g|\right)^{1/3}\,, \\
  C'_{t12} &= - \frac{C_{t12}}{A^2 \left(C_{t12}^2 - |g|\right)}\,.
 \end{split}
\end{equation}

\section{Timelike SL(5) duality} \label{Sd4}
We saw in section \ref{STIntSym} that there are four different types of generalised metric that one ought to consider. These differ in the signature of the block-diagonal components of the generalised metric, $\gm_{ij}$ and $\gm^{ij,kl}$. In the four-dimensional case the generalised metric of type I is given by
\begin{equation}
 \gm_{MN} = |g_{11}|^{-1/2} \left( \begin{array}{ccc}
                            g_{ik} \left[ \delta^k_j \left(1 - V^2\right) + V^k V_j \right] & \frac{1}{\sqrt{2}} C_i^{\ph{i}mn} & 0 \\
                            \frac{1}{\sqrt{2}} C^{kl}_{\ph{kl}j} & g^{k[m}g^{n]l} & 0 \\
                            0 & 0 & g_{AB} \end{array} \right)
\end{equation}
in the $\left(g_{11}, C_3\right)$ frame and by
\begin{equation}
 \gm_{MN} = |\og_{11}|^{-1/2} \left( \begin{array}{ccc}
                            \og_{ij} & \frac{1}{\sqrt{2}} \Omega_{i}^{\ph{i}mn} & 0 \\
                            \frac{1}{\sqrt{2}} \Omega^{kl}_{\ph{kl}j} & g^{kp}g^{ql} \left[ \delta_{[pq]}^{mn} \left(1 - W^2\right) + \delta_{[p}^m W_{q]} W^n \right] & 0 \\
                            0 & 0 & g_{AB} \end{array} \right)
\end{equation}
in the $\left(\og_{11}, \Omega_3\right)$ frame. We define
\begin{align}
 C_{ijk} &= \epsilon_{ijkl}V^l\,, \\
 \Omega^{ijk} &= \bar{\epsilon}^{ijkl} W_l\,, \\
 V^2 &= V^i V^j g_{ij}\,, \\
 W^2 &= W_i W_j \og^{ij}\,,
\end{align}
where $\epsilon_{ijkl}$ and $\bar{\epsilon}^{ijkl}$ are the components of the Levi-Civita tensors for $g$ and $\og$, respectively. By analogy with the three-dimensional case discussed in section \ref{STIntSym} we study the eigenvectors of the matrix
\begin{equation}
 \delta^i_j \left(1- V^2\right) + V^i V_j\,,
\end{equation}
and find that it has eigenvalues $\lambda = 1$ of multiplicity one, corresponding to eigenvectors parallel to $V^i$ and $\lambda = 1-V^2$ of multiplicity three for eigenvectors perpendicular to $V^i$. Thus, it can either have four positive eigenvalues when $V^2 < 1$ or one positive and three negative (or zero) eigenvalues when $V^2 \geq 1$ with the singularity occurring when $V^2 = 1$. The generalised metric of type I has signatures $\gm_{ij} = \left(-,+,+,+\right)$ and $\gm^{ij,kl} = \left(-,-,-,+,+,+\right)$ for these components and thus $V^2 < 1$. To simplify the notation we will denote the signatures by $\left(p,q\right)$ where $p$ denotes the number of timelike and $q$ the number of spacelike directions. Thus, when $V^2 < 1$, $\gm_{ij}$ has the same signature as $g_{ij}$, i.e. $\left(1,3\right)$. When $V^2 \geq 1$, on the other hand, $\gm_{ij}$ has the opposite signature for the three directions perpendicular to $V^i$. Because $V^2 \geq 1$, these always include time and two spatial directions, thus giving signature $\left(2,2\right)$. $\gm_{ij}$ will never have signature $\left(4,0\right)$. Similar arguments can be applied to the $\gm_{ij,kl}$ components to show that it could have signature $\left(3,3\right), \left(4,2\right)$ or be singular. We see that the generalised metric will again be of four types as summarised in table \ref{T4DFrameConditions} and we see that we have similar structures as in the three-dimensional case. For example, we need to check that the generalised metric component $\gm_{ij}$ has not changed the sign along time and two spacelike directions. The only complication arises because one may have various non-zero components of $C_3$ and $\Omega_3$. However, the ``building blocks'' are the same as for three dimensions. This should not be a surprise: this structure is due to the 3-form and trivector which have three components.
\begin{table}
\centering
\begin{tabular}{|c|c|c||c|}
\hline
Type & $\gm_{ij}$ & $\gm^{ij,kl}$ & Minimal valid frames \\
\hline
I & $\left(1,3\right)$ & $\left(3,3\right)$ & $\left(g_{11}, C_3\right)$ or $\left(\og_{11}, \Omega_3\right)$ \\
II & $\left(2,2\right)$ or $0$ & $\left(3,3\right)$ & $\left(g_{11}, C_3\right)$ \\
III & $\left(1,3\right)$ & $\left(4,2\right)$ or $0$ & $\left(\og_{11}, \Omega_3\right)$ \\
IV & $\left(2,2\right)$ or $0$ & $\left(4,2\right)$ or $0$ & $\left(\hat{g}_{11}, C_3, \Omega_3\right)$ \\
\hline
\end{tabular}
\caption{The conditions for being able to use a certain field frame $\left(g_{11},C_3\right), \left(\og_{11},\Omega_3\right), \left(\hat{g}_{11},C_3,\Omega_3\right)$ in terms of the signature of the components of the generalised metric $\gm_{ij}$ and $\gm^{ij,kl}$. The signature of $p$ timelike and $q$ spacelike directions is denoted by $\left(p,q\right)$. The minimal valid frames are those with the smallest number of bosonic fields that describe the physics. $\left(\hat{g}_{11},C_3,\Omega_3\right)$ is always a valid frame but we only include it when it is the only valid frame because it otherwise carries an unnecessary redundancy.}
\label{T4DFrameConditions}
\end{table}

\subsection{Transformation laws under $U_\Omega$}
We will now give the transformation law for the metric and 3-form under the $U_\Omega$ transformation where $\Omega^{ijk} = \epsilon^{ijkz} A \sqrt{|g|}$. $z$ is a placeholder labelling either a spacelike or a timelike direction. We find
\begin{equation} \label{ESL5Trans}
 \begin{split}
  g'_{ij} &= \left[ g_{ij} - A \sqrt{|g_4|} \left(  V_i \delta_i^z + V_j \delta_i^z \right) - A^2 |g_4| \left(1-V^2\right) \delta_i^z \delta_j^z \right] \left[ \left(1-AV^z \sqrt{|g_4|}\right)^2 - A^2 |g_4| g^{zz} \right]^{-2/3}\,, \\
  C'_{ijk} &= \frac{C_{ijk} \left(1-A\sqrt{|g_4|}V^z\right) + A \sqrt{|g_4|} \epsilon_{ijkl} g^{lz}}{\left(1-A\sqrt{|g_4|}V^z\right)^2 - A^2 |g_4| g^{zz}}\,.
 \end{split}
\end{equation}
This transformation law is valid when the function $f = \left(1-A\sqrt{|g_4|}V^z\right)^2 - A^2 |g_4| g^{zz}$ is positive definite. When this does not hold we must include a trivector as for $SL(2) \times SL(3)$. Note that when $z$ is timelike, $g^{zz} < 0$ and so $f > 0$ is always satisfied. This should not be surprising because we are performing the duality along spacelike directions and so we can always gauge away the trivector.

For a diagonal metric with only one non-zero component of $V$ labelled by $V^w$ we split the equations as $x^i = \left( x^\alpha, w, z \right)$ so that the transformed fields simplify to
\begin{equation}
  \begin{split}
   ds'^2 &= ds_{\alpha\beta}^2 \left(1 - A^2 |g_4| g^{zz}\right)^{-2/3} + g_{ww} \left(dw - A \sqrt{|g_4|} V^w dz\right)^2 \left(1 - A^2|g_4| g^{zz}\right)^{-2/3} \\
   & \quad + g_{zz} dz^2 \left(1 - A^2 |g_4| g^{zz} \right)^{1/3}\,, \\
   C'_{z\alpha\beta} &= \frac{C_{z\alpha\beta}}{1 - A^2 |g_4| g^{zz}}\,, \\
   C'_{\alpha\beta w} &= \frac{-A|g_4| \eta_{\alpha\beta w z}g^{zz}}{1 - A^2 |g_4| g^{zz}}
  \end{split}
\end{equation}
and all other components vanishing. $\eta_{\alpha \beta w z}$ is the alternating symbol where $\eta_{12wz} = 1$, etc.

We will see how this can be used to generate momentum in section \ref{SESL5}.

\section{Examples} \label{SExamples}
In \cite{Malek:2012pw} we dualised specific examples of Euclidean 11-dimensional supergravity. We had a glimpse at dualities acting in timelike directions by taking a Lorentzian solution but first Wick-rotating to Euclidean 11-dimensional supergravity, then dualising and finally Wick-rotating back. We found that this naive procedure can cause difficulties. For example, a Buscher duality of the extreme M2-brane seems to give rise to a singular solution, while the $U_\Omega$ transformation acting on uncharged black M2-branes gives a black M2-brane like solution but with harmonic functions that may be negative. The spacetime metric is then complex. We now revisit these examples and find that these problems arose because we were using the $\left(g_{11}, C_3\right)$ frame even when it was not valid. Using the right frame, $\left(\og_{11}, \Omega_3\right)$, $\left(\hg_{11}, \hat{C}_3, \hat{\Omega}_3\right)$ or $\left(\cg_{11}, \check{C}_3, \check{\Omega}_3\right)$, as listed in tables \ref{TFrameConditions} and \ref{T4DFrameConditions}, we find well-behaved dual solutions instead.

From equation \eqref{ECT} we see that under $U_\Omega$, the coordinates transform
\begin{equation}
 \begin{split}
  x^i &\rightarrow x^i - \frac{1}{2} \Omega^{ijk} y_{jk}\,, \\
  y_{ij} &\rightarrow y_{ij}\,, \\
  x^A & \rightarrow x^A\,.
 \end{split}
\end{equation}
Thus, if we start with a conventional solution of the generalised Lagrangian \eqref{EEGL}, i.e. having no dependence on the dual coordinates $y_{ij}$, the transformed solution will be independent of the dual coordinates $y_{ij}$ as long as $\Omega^{ijk}$ has non-zero components along isometries only. Explicitly
\begin{equation}
 \begin{split}
  \partial_{i} & \rightarrow \partial_i\,, \\
  \partial^{ij} & \rightarrow \partial^{ij} + \Omega^{ijk} \partial_k\,,
 \end{split}
\end{equation}
and we see that we preserve the sectioning condition $\partial^{ij} = 0$ if we act with $\Omega^{ijk}$ along isometries only. The supergravity solutions corresponding to M2-branes are then natural examples to consider since they contain three isometries, corresponding to the worldvolume directions. We begin by acting with $SL(2) \times SL(3)$ along the worldvolume directions on uncharged and extreme M2-brane solutions before studying the action of $SL(5)$ in section \ref{SESL5}.

\subsection{Uncharged black M2-brane} \label{SUncharged}
We begin with the example of an uncharged black M2-brane \cite{Gueven:1992hh}.
\begin{equation}
 \begin{split}
  ds^2 &= -W dt^{2} + dy_1^2 + dy_2^2 + W^{-1} dr^2 + r^2 d\Omega_{(7)}^{2}\,, \\
  W &= 1 + h/r^{6}\,, \\
  C_{t12} &= 0\,,\footnotemark
 \end{split}
\end{equation}
\footnotetext{The indices $1,2$ correspond to $y_1, y_2$, respectively.}where $r$ is the radius in the six transverse directions, $d\Omega_{(7)}^{2}$\footnote{The symbol $\Omega$ is used here for two different purposes: once in relation to a $S^7$ and once for the trivector. The context will make it clear what is being meant.} corresponds to the metric of a $S^7$ and $\omega_d$ is the volume of a $S^d$
\begin{equation}
 \omega_d = \frac{2\pi^{\frac{d+1}{2}}}{\Gamma(\frac{d+1}{2})}\,.
\end{equation}

The tension of the brane is
\begin{equation}
 M_2 = -\frac{9 h \omega_7}{2\kappa^2}\,,
\end{equation}
where $\kappa^2 = 8\pi G_N^{(11)}$ and $G_N^{(11)}$ is the 11-dimensional Newton's constant. The tension is positive for $h < 0$ and we will write
\begin{equation}
 -h = k > 0\,.
\end{equation}

We want to act with the three-dimensional U-duality group $E_3$ along the three world-volume isometries. However, we know that there are only two families of non-trivial transformations, generated by $U_\Omega$ and $U_B$.\footnote{By trivial we mean those dualities acting as gauge transformations, i.e. either rigid diffeomorphisms and 3-form shifts.}

\subsubsection{$U_{\Omega}$ acting on uncharged black M2-brane} \label{SUM2Omega}
We first consider $U_\Omega$ where $\Omega^{t12} = A$. We saw in section \ref{STIntSym} that depending on the sign of
\begin{equation}
 \begin{split}
  f &\equiv 1 - A^2|g| \\
  &= 1 - A^2 W \\
  &= 1 - A^2 + \frac{A^2k}{r^6}\,,
 \end{split}
\end{equation}
we may need to include the trivector. Because there is no initial 3-form, the dual generalised metric is of type I if $f > 0$ and type III if $f \leq 0$. We consider three cases $A^2 < 1$, $A^2 = 1$ and $A^2 > 1$.

\paragraph{Case 1: $A^2 < 1$.} In this case $f$ is positive everywhere and we can describe the solution in the $\left(g_{11},C_3\right)$ frame. We rescale the coordinates in order to obtain an asymptotically flat solution\footnote{Throughout this section this will mean asymptotically flat with respect to the transverse coordinates when we say "asymptotically flat".}
\begin{equation}
 \begin{split}
  t & \rightarrow T = t (1-A^2)^{-1/3}\,, \\
  y_1 & \rightarrow Y_1 = y_1 (1-A^2)^{-1/3}\,, \\
  y_2 & \rightarrow Y_2 = y_2 (1-A^2)^{-1/3}\,, \\
  r & \rightarrow R = r (1-A^2)^{1/6}\,.
 \end{split}
\end{equation}
and obtain
\begin{equation}
 \begin{split}
  ds'^2 &= G^{-2/3} \left(- W dT^2 + dY_1^2 + dY_2^2 \right) + G^{1/3} \left(W^{-1} dR^2 + R^2 d\Omega_{(7)}^{2}\right)\,,\\
  C'_{T Y_1 Y_2} &= - \frac{1}{A} G^{-1} + \textrm{const.} \label{Harrison}\,,
 \end{split}
\end{equation}
where now
\begin{equation}
 \begin{split}
  G &= 1 + \frac{A^2 k}{R^6}\,,\\
  W &= 1 - \frac{k (1 - A^2)}{R^6}\,. \label{HarrisonHarmonics}
 \end{split}
\end{equation}
We found this dual solution in \cite{Malek:2012pw} by Wick rotating before and after applying spacelike dualities, and noted that it is the solution of a charged M2-brane of tension and charge density
\begin{equation}
 \begin{split}
  M_2 &\rightarrow M'_2 = \left(1 - \frac{1}{3}A^2 \right) M_2\,, \\
  Q &\rightarrow Q' = - \frac{2}{3} A M_2 = - \frac{2 A M'_2}{3 - A^2}\,.
 \end{split}
\end{equation}
We see that if $A^2 < 1$ the $U_\Omega$ transformation charges the brane solution. This is thus a generalisation of the Harrison transformation of Einstein-Maxwell theory \cite{Harrison:1968,Breitenlohner:1987dg}.

\paragraph{Case 2: $A^2 = 1$.} This transformation belongs to the quantum group $E_3(Z)$ where $A \in Z$ is an integer.
\begin{equation}
 f = \frac{k}{r^6}
\end{equation}
is again positive everywhere so that we can use the $\left(g_{11}, C_3\right)$ frame. We find the dual solution
\begin{equation}
 \begin{split}
  ds'^2 &= f^{-2/3}\left(-W dt^{2} + dy_1^2 + dy_2^2\right) + f^{1/3} \left(W^{-1} dr^2 + r^2 d\Omega_{(7)}^{2}\right)\,, \\
  C'_{t12} &= \frac{r^{6}}{k} - 1\,.
 \end{split}
\end{equation}
Upon changing coordinates to $\rho = \frac{k^{-1/3}}{2} r^2$ we recognize this as a Schwarzschild-$AdS_4 \times S^7$ solution
\begin{align}
 ds'^2 &= \left(\frac{\rho}{R}\right)^2 \left[ - \left(1 - \frac{\sqrt{k}}{8 \rho^3} \right) dt^2 + dy_1^2 + dy_2^2 \right] + \left(\frac{R}{\rho}\right)^2 \frac{1}{\left(1- \frac{\sqrt{k}}{8\rho^3}\right)} d\rho^2 + k^{1/3} d\Omega_{(7)}^2\,, \\
 C'_{t12} &= \frac{\rho^3}{k} - 1\,,
\end{align}
where $R = \frac{k^{1/6}}{2}$ and the field strength of the 3-form gives the cosmological constant for the Schwarzschild-$AdS_4$ part and its dual gives the volume form of the $S^7$. This can be viewed as the 11-dimensional analogue of a ``subtracted geometry'' solution which can be constructed by removing the asymptotically flat region of the original solution \cite{Cvetic:2012tr}. The subtracted geometry of a specific intersecting brane solution that gives rise to the Kerr-Newman black hole upon compactification to four dimensions has recently been shown to lie in the orbit of Harrison transformations acting on the initial solution \cite{Virmani:2012kw}.\footnote{Subtracted geometries of four-dimensional Kerr-Newman black holes manifestly exhibit the ``hidden'' conformal symmetry of the black hole solutions \cite{Castro:2010fd,Cvetic:2011hp,Cvetic:2011dn,Cvetic:2012tr,Compere:2010uk,Bertini:2011ga}. These conformal symmetries are important for the Kerr/CFT correspondence. For a comprehensive review of the Kerr/CFT correspondence, see \cite{Compere:2012jk}. They are also useful because the scalar wave equation becomes separable. Thermodynamic quantities, which remain invariant under the ``subtraction'', can then be computed with ease.}

\paragraph{Case 3: $A^2 > 1$.} The function $f$ is positive only close to the brane when
\begin{equation}
 r < \left(\frac{A^2k}{A^2 - 1}\right)^{1/6}\,,
\end{equation}
and so in this region we can describe the solution in the $\left(g_{11},C_3\right)$ frame to obtain
\begin{equation}
 \begin{split}
  ds'^2 &= G^{-2/3} \left(- W dT^2 + dY_1^2 + dY_2^2 \right) + G^{1/3} \left(W^{-1} dR^2 + R^2 d\Omega_{(7)}^{2}\right)\,,\\
  C'_{T Y_1 Y_2} &= A + \frac{1}{A} \left(G^{-1} - 1\right)\,,
 \end{split}
\end{equation}
where now
\begin{equation}
 \begin{split}
  G &= -A^2 + 1 + \frac{A^2 k}{r^6}\,,\\
  W &= 1 - \frac{A^2k}{r^6}\,.
 \end{split}
\end{equation}
We see this solution causes problems only where $r \geq \left(\frac{A^2k}{A^2-1}\right)^{1/6}$ which is where it is not valid. Because the solution is not valid globally, we cannot describe its charge or mass through the Komar procedure.

Alternatively, we can construct a global dual solution in the $\left(\og_{11}, \Omega_3\right)$ frame since the generalised metric is of type III. We saw in section \ref{STU} that the $U_\Omega$ transformation just shifts $\Omega^{ijk}$
\begin{equation}
 \Omega^{t12} \rightarrow \Omega^{t12} + A\,.
\end{equation}
As a result the dual fields in the $\left(\og_{11},\Omega_3\right)$ frame are
\begin{equation} \label{EM2Trivector}
 \begin{split}
  d\bar{s}'^2 &= -W dt^2 + dy_1^2 + dy_2^2 + W^{-1} dr^2 + r^2 d\Omega_{(7)}^{2}\,, \\
  \Omega'^{t12} &= A\,.
 \end{split}
\end{equation}
The solutions \eqref{Harrison} fit into a one-parameter family of charged and uncharged non-extremal black branes, including solutions corresponding to negative mass when $A^2 > 3$. If we had naively used the timelike transformation rules as in equation \eqref{ETOmega}, without checking that we are using the right frame, we would have obtained these negative-mass solutions. This is what happened in \cite{Malek:2012pw} but here we see that the family of dual solutions contains the uncharged black M2-brane corresponding to $A = 0$, charged black ones obtained by the Harrison transformation when $A^2 < 1$, the subtracted geometry solution for $A^2 = 1$ and finally dual solutions including a trivector, given by equation \eqref{EM2Trivector}, when $A^2 > 1$. The duality orbit avoids the unphysical solutions thanks to the trivector.

In \cite{Virmani:2012kw} the subtracted geometry of a different spacetime is generated by Harrison transformations. There the dualities are used as a solution-generating mechanism on dimensionally reduced spaces. Our result confirms this finding for a much simpler example but does so directly at the level of the 11-dimensional solutions without the need to dimensionally reduce. We can thus see that the subtracted geometry of a brane configuration can be generated by the $U_\Omega$ transformation which is a generalisation of the Harrison transformation. Furthermore, we see by comparison that here the value $A^2 = 1$ corresponds to an ``infinite'' Harrison boost. Thus, the transformations for $A^2 > 1$ do not arise in the conventional picture of dualities. In the context of generalised geometry, on the other hand, there is no reason to cut off the parameter at $A^2 = 1$ except that dualities for $A^2 > 1$ will include a non-zero trivector and thus go beyond the conventional description of 11-dimensional supergravity.

\subsubsection{Buscher duality of uncharged M2-brane} \label{SUM2Buscher}
We can go through the same procedure when acting with $U_B$
\begin{equation}
 U_B = \left( \begin{array}{cc}
                0 & \frac{1}{\sqrt{2}} D \\
                \frac{1}{\sqrt{2}} K & 0
                \end{array} \right)\,,
\end{equation}
where $D^{t12} = A$ and $K_{t12} = -\frac{1}{A}$. The dual generalised metric
\begin{equation}
 \gm' = |g_{11}|^{-1/2}
        \left( \begin{array}{cc}
                -A^{-2} |g|^{-1} g & 0 \\
                0 & -A^2 |g| g^{-1} g^{-1}
                \end{array} \right)
\end{equation}
is of type IV and thus can only be interpreted using both the 3-form and trivector. Using equations \eqref{ETBuscherCheck} and \eqref{ETBuscherHat} we obtain the fields in the checked frame
\begin{equation}
 \begin{split}
  d\check{s}^2 &= \left(\frac{\sinh\theta}{A \sqrt{W}}\right)^{4/3} \left(- W dt^2 + dy_1^2 + dy_2^2 \right) + \left(\frac{\sinh\theta}{A \sqrt{W}}\right)^{-2/3} \left(W^{-1} dr^2 + r^2 d\Omega_{(7)}^2 \right)\,, \\
  \check{C}_{t12} &= - \frac{\sinh2\theta}{2 A^2 \sqrt{W}}\,, \\
  \check{\Omega}^{t12} &= A^2 \sqrt{W} \coth \theta\,,
 \end{split}
\end{equation}
and in the hatted frame
\begin{equation}
 \begin{split}
  d\hat{s}^2 &= \left(A \sqrt{W} \sinh\theta\right)^{-4/3} \left(- W dt^2 + dy_1^2 + dy_2^2 \right) + \left(A \sqrt{W} \sinh\theta\right)^{2/3} \left(W^{-1} dr^2 + r^2 d\Omega_{(7)}^2 \right)\,, \\
  \hat{C}_{t12} &= - \left(A^2 \sqrt{W} \coth\theta\right)^{-1}\,, \\
  \hat{\Omega}^{t12} &= \frac{A^2 \sqrt{W}\sinh2\theta}{2}\,.
 \end{split}
\end{equation}
The concepts of mass and charge are not well-defined here because of the appearance of the trivector. Furthermore, because the parameter $\theta$ can be chosen locally, we may be better off using the generalised metric instead of this decomposition in terms of the metric, 3-form and trivector.

Once again, if we had used the timelike Buscher rules \eqref{ETBuscher} naively without checking the validity of the $\left(g_{11},C_3\right)$ frame we would have obtained a solution corresponding to a negative mass. But the trivector saves us so that we do not get ``unphysical'' dual solutions.

\subsection{Extreme M2-brane} \label{SEC}
We will now repeat the analysis for the extreme M2-brane \cite{Duff:1990xz} with the following coordinate and gauge choice
\begin{equation}
 \begin{split}
  ds^2 &= H^{-2/3} \left(-dt^2 + dy_1^2 + dy_2^2\right) + H^{1/3} \left(dr^2 + r^2 d\Omega_{(7)}^{2}\right)\,, \\
  C_{t12} & = H^{-1} + n\,, \\
  H &= 1 + \frac{h}{r^6}\,.
 \end{split}
\end{equation}

\subsubsection{$U_\Omega$ acting on extreme M2-brane} \label{SECOmega}
Acing with $U_\Omega$, where $\Omega^{t12} = A$, we have to study the sign of the function
\begin{equation}
 f = \left(1 + An\right)^2 + 2AH^{-1}\,.
\end{equation}
Because $V^2 = 1$, the dual generalised metric is of type II when $f>0$ and thus expressible in the $\left(\og_{11}, C_3\right)$. If, on the other hand, $f \leq 0$ we have to include the trivector and will obtain a family of dual solutions, linked by local $SO(1,1)$ rotations, as before.

\paragraph{Case 1: $\left(1 + An\right)^2 + 2A > 0$.} Now $f > 0$ everywhere so we can remove the trivector field by a gauge transformation. We then obtain the dual spacetime from equation \eqref{ETOmega}
\begin{equation} \label{EOmegaExtreme3}
\begin{split}
ds'^2 &= G^{-2/3} \left(-dT^2 + dY_1^2 + dY_2^2\right) + G^{1/3} \left(dR^2 +
R^2 d\Omega_{(7)}^2\right)\,, \\
C'_{TY_1Y_2} &= G^{-1} + \textrm{const.}\,, \\
G &= 1 + \left(1+An\right)^2 \frac{h}{R^6}\,.
\end{split}
\end{equation} 
Here the coordinates $T, Y_1, Y_2, R$ are chosen to make the solution
asymptotically flat. This is the same result as obtained by Wick-rotations in
\cite{Malek:2012pw} and corresponds to a new extreme M2-brane of different
tension and charge. These are given by
\begin{equation}
 \begin{split}
  M'_2 &= M_2 \left(1 + An \right)^2\,, \\
  Q' &= Q \left( 1 + An \right)^2\,,
 \end{split}
\end{equation}
and we see that if we use the quantum U-duality group $E_3(Z)$ so that $A, n \in Z$ are integers, we obtain a dual extreme M2-brane with tension and charge that are multiples of the old ones. Thus, mass and charge quantisation would be preserved by the discrete quantum duality group. We also notice that there is a large degeneracy amongst the solutions we generate: while we have two free parameters in the duality $A, n$, the dual solutions depend only on the combination $An$. Thus, if $n = 0$ we always obtain the same extreme M2-brane as the one we started with just as we found in \cite{Malek:2012pw}.

\paragraph{Case 2: $\left(1+An\right)^2 + 2A = 0$.} We still have $f > 0$ everywhere so that we can use the $\left(g_{11}, C_3\right)$ frame but we cannot make the solution asymptotically flat. This is analogous to the case of the uncharged black M2-brane in section \ref{SUM2Omega}. We obtain the solution corresponding to the ``subtracted'' geometry \cite{Cvetic:2012tr} which is given by
\begin{equation}
 \begin{split}
  ds'^2 &= \left[ \left(1 + An\right)^2 \frac{h}{r^6} \right]^{-2/3} \left(-dt^2 + dy_1^2 + dy_2^2 \right) + \left[ \left(1 + An\right)^2 \frac{h}{r^6} \right]^{1/3} \left(dr^2 + r^2 d\Omega_{(7)}^2 \right)\,, \\
  C'_{t12} &= \frac{1 - \left(1+An\right)An^2}{\left(1+An\right)^2 h} r^6 + \frac{n}{1+An}\,.
 \end{split}
\end{equation}
Here the subtracted geometry is just the near-horizon limit of the M2-brane, $AdS_4 \times S^7$, because there is only one harmonic function, in contrast to the uncharged case, so subtracting the asymptotically flat region will give the near-horizon limit.

\paragraph{Case 3: $\left(1+An\right)^2 + 2A < 0$.} In this case $f \leq 0$ in some regions and the dual generalised metric is of type IV. Thus we need to include the trivector and the checked and hatted frame fields can be calculated from equations \eqref{ETOmegaCheck} and \eqref{ETOmegaHat}.
\begin{equation}
 \begin{split}
  d\check{s}'^2 &= \left( \frac{e^\theta}{\sqrt{H}} + n \sqrt{H} \sinh\theta \right)^{4/3} \left(-dt^2 + dy_1^2 + dy_2^2\right) \\
  & \quad + \left(  \frac{e^\theta}{\sqrt{H}} + n \sqrt{H} \sinh\theta \right)^{-2/3} \left(dr^2 + r^2 d\Omega_{(7)}^2 \right)\,, \\
  \check{C}'_{t12} &= \left( \frac{e^\theta}{\sqrt{H}} + n \sqrt{H} \sinh\theta\right) \left( \frac{e^\theta}{\sqrt{H}} + n\sqrt{H} \cosh\theta \right)\,, \\
  \check{\Omega}'^{t12} &= \frac{A H^{-1} e^\theta + \sinh\theta \left(1 + An \right)}{H^{-1} e^\theta + n\sinh\theta}\,,
 \end{split}
\end{equation}
and
\begin{equation}
 \begin{split}
  d\hat{s}'^2 &= \left[ A \frac{e^\theta}{\sqrt{H}} + \cosh\theta \sqrt{H} \left(1 + An\right) \right]^{-4/3} \left(-dt^2 + dy_1^2 + dy_2^2 \right) \\
  & \quad + \left[ A \frac{e^\theta}{\sqrt{H}} + \cosh\theta \sqrt{H} \left(1 + An\right) \right]^{2/3} \left(dr^2 + r^2 d\Omega_{(7)}^2 \right)\,, \\
  \hat{C}'_{t12} &= \frac{H^{-1} e^\theta + n\cosh\theta}{A H^{-1} e^\theta + \cosh\theta \left(1+An\right)}\,, \\
  \hat{\Omega}'^{t12} &= \left[ A \frac{e^\theta}{\sqrt{H}} + \sqrt{H} \sinh\theta \left(1 + An\right) \right] \left[ A \frac{e^\theta}{\sqrt{H}} + \sqrt{H} \cosh\theta \left(1 +An\right) \right]\,.
 \end{split}
\end{equation}

\subsubsection{Buscher duality of extreme M2-brane} \label{SECBuscher}
If we act with a Buscher duality $U_B$, the dual generalised metric is given in equation \eqref{EGMBuscher}
\begin{equation}
 \gm' = \left(-1\right)|g_{11}|^{-1/2}
	\left( \begin{array}{cc}
        A^{-2}|g|^{-1} g & \frac{1}{\sqrt{2}} C g^{-1} g^{-1} \\
	\frac{1}{\sqrt{2}} g^{-1} g^{-1} C & A^2 |g| \left(1-V^2\right) g^{-1} g^{-1}
        \end{array} \right)\,.
\end{equation}
\paragraph{Case 1 $n > 0$.}
\begin{equation}
 V^2 = \left(1 + n H\right)^2 > 1
\end{equation}
everywhere if $n > 0$. We can then describe the dual solution in the $\left(g_{11}, C_3\right)$ frame. Using equation \eqref{ETBuscher} and rescaling the coordinates to make the solution asymptotically flat we find
\begin{equation}
 \begin{split}
  ds'^2 &= G^{-2/3} \left(-dt^2 + dy_1^2 + dy_2^2\right) + G^{1/3} \left(dr^2 + r^2 d\Omega_{(7)}^2\right)\,, \\
  C'_{t12} &= G^{-1} - 2 - n\,, \\
  G &= 1 + A^2n^2 \frac{h}{R^6}\,.
 \end{split}
\end{equation}
We see that this is once again an extreme M2-brane with different tension and charge proportional to the initial ones
\begin{equation}
 \begin{split}
  M'_2 &= A^2 n^2 M_2\,, \\
  Q' &= A^2 n^2 Q\,.
 \end{split}
\end{equation}
Again this means that if we use the quantum U-duality group $E_3(Z)$ and $A, n \in Z$ are integers, mass and charge remain quantised appropriately.

\paragraph{Case 2: $n \leq 0$.} This now means that we have to use a trivector field in the checked or hatted frames. We can find the fields using equations \eqref{ETBuscherCheck} and \eqref{ETBuscherHat} but we will omit them.

\subsection{$SL(5)$ and generating momentum} \label{SESL5}
We want to act with $SL(5)$ on brane-like solutions but in a way that does not simply reduce to $SL(2) \times SL(3)$. In order to achieve this, we perform a gauge transformation to have two non-zero components of $C_3$. We start with the seed solution
\begin{equation}
 \begin{split}
  ds^2 &= H^{-2/3} \left(-dt^2 + dy_1^2 + dy_2^2\right) + H^{1/3} \left(dx_3^2 + dz_4^2 + \ldots + dz_{10}^2\right)\,, \\
  C_{t12} &= H^{-1} + n\,, \\
  C_{123} &= k\,.
 \end{split}
\end{equation}
We first consider acting with $\Omega^{t12}$ so that we can take $H = 1 + \frac{h}{r^6}$ and $r^2 = x_3^2 + z_4^2 + \ldots + z_{10}^2$.\footnote{The convention of labelling the transverse coordinates by $\left(x_3, z_4 \ldots z_{10}\right)$ has been chosen to facilitate the discussion of the smeared M2-brane.}

Because the duality acts along $t, y_1, y_2$ only, we can perform a $x_3$-dependent gauge transformation on $C_{123}$ before dualising so that $k = k(x_3)$ in general. In this case, the resultant solution is
\begin{equation}
\begin{split}
ds'^2 &= \left[ -\left(dt - A k(x_3) dx_3\right)^2 + \overrightarrow{dy}^2
\right] j^{-2/3} + \left[ dx_3^2 + \overrightarrow{dz}^2 \right] j^{1/3}\,,\\
C'_{t12} &= \frac{1+2An+\left(1+An\right)nH}{j}\,, \\
C'_{123} &= f(x_3) \frac{A + \left(1+An\right)H}{j}\,,
\end{split}
\end{equation} 
where $j = H \left(1+An\right)^2 + 2A \left(1+An\right)$. By using a different
coordinate frame
\begin{equation}
\begin{split}
T &= \left[\left(1+An\right)\left(1+An+2A\right)\right]^{-1/3} \left( t - A \int
f(x_3) dx_3 \right)\,, \\
\overrightarrow{Y} &= \overrightarrow{y}
\left[\left(1+An\right)\left(1+An+2A\right)\right]^{-1/3}\,, \\
X_3 &= x_3 \left[\left(1+An\right)\left(1+An+2A\right)\right]^{1/6}\,, \\
\overrightarrow{Z} &= \overrightarrow{z}
\left[\left(1+An\right)\left(1+An+2A\right)\right]^{1/6}\,, \\
R &= r \left[\left(1+An\right)\left(1+An+2A\right)\right]^{1/6}\,,
\end{split}
\end{equation}
we see that the solution corresponds to another extreme M2-brane:
\begin{equation}
\begin{split}
ds'^2 &= \left[ -dT^2 + \overrightarrow{dY}^2 \right] G^{-2/3} + \left[ dX_3^2 +
\overrightarrow{dZ}^2 \right] G^{1/3}\,, \\
C'_{TY_1Y_2} &= \frac{1}{G} + \textrm{const.}\,, \\
C'_{Y_1Y_2X_3} &= 0 \,,
\end{split}
\end{equation}
where
\begin{equation}
 G = 1 + \left(1+An\right)^2 \frac{h}{R^6}\,.
\end{equation}
Thus, the resultant tension and charge are mutliples of the initial ones
\begin{equation}
\begin{split}
M' &= M \left(1+An\right)^2\,, \\
Q' &= Q \left(1+An\right)^2\,,
\end{split}
\end{equation}
giving the same results as for $k = 0$.

We can also smear the brane in the $x_3$ direction so that $H = 1 + \frac{h}{r^5}$ where $r^2 = z_4^2 + \ldots + z_{10}^2$. Now, there are four isometries $t, y_1, y_2, x_3$ and we can also act with $\Omega^{123}$. This will give a dual spacetime metric with off-diagonal components due to $C_{t12}$. Using equations \eqref{ESL5Trans} we get
\begin{equation}
 \begin{split}
  ds'^2 &= \left( g_{tt} dt^2 + g_{AB} dx^A dx^B \right) \left( (1 - A C_{123})^2 - A^2 |g_4| g^{tt} \right)^{1/3} \\
  & \quad + \left( g_{33} \left( dx_3 + A C_{t12} dt \right)^2 + g_{\alpha\beta} dx^\alpha dx^\beta \right) \left( (1 - A C_{123})^2 - A^2 |g_4| g^{tt} \right)^{-2/3} \,,
 \end{split}
\end{equation}
and the transformed 3-forms are
\begin{equation}
\begin{split}
C'_{t12} &= \frac{C_{t12} \left(1 - A C_{123} \right)}{\left(1 - A C_{123}
\right)^2 - A^2 |g_4| g^{tt}}\,, \\
C'_{123} &= \frac{C_{123} \left(1 - A C_{123} \right) + A |g_4| g^{tt}}{ \left(1
- A C_{123} \right)^2 - A^2 |g_4| g^{tt}}\,.
\end{split}
\end{equation}
Here the indices $A,B$ still label the transverse undualisable directions but
$\alpha, \beta$ label dualisable directions other than $t$ and $x_3$. In the
case at hand $x^{\alpha} = \left(y_1, y_2\right)$ and $x^A = \left(z_4, \ldots
z_{10}\right) \equiv \overrightarrow{z}$.
This expression can be evaluated to be
\begin{equation}
\begin{split}
ds'^2 &= \left[ \left(1-Ak\right)^2 H + A^2 \right]^{1/3} \left(d\mathbf{z}^2 -
H^{-1} dt^2\right) \\
& \quad + \left[ \left(1-Ak\right)^2 H + A^2 \right]^{-2/3} \left(d\mathbf{y}^2
+ H\left(dx_3 + A H^{-1} (1+nH) dt\right)^2 \right) \,, \\
C'_{t12} &= \frac{\left(1+nH\right)\left(1-Ak\right)}{\left(1-Ak\right)^2 H +
A^2}\,, \\
C'_{123} &= \frac{-A + kH \left(1-Ak\right)}{\left(1-Ak\right)^2 H +A^2}\,,
\end{split}
\end{equation}
where $\mathbf{y}^2 = y_1^2 + y_2^2$ and $\mathbf{z}^2 = z_4^2 + \ldots +
z_{10}^2$.
By an appropriate coordinate transformation we write the solution in the more suggestive form
\begin{equation} \label{EAnsRot1}
\begin{split}
ds'^2 &= P^{-2/3} \left\{ -Q \left[dt - L dx_3 \right]^2 + d\mathbf{y}^2
\right\} \\
& \quad + P^{1/3} \left\{ d\mathbf{z}^2 + Q^{-1} dx_3^2 \right\}\,, \\
F'_{ry_1y_2x_3} &= - \beta P^{-2} \partial_r P\,, \\
F'_{try_1y_2} &= - \alpha P^{-2} \partial_r P\,.
\end{split}
\end{equation}
\paragraph{Case 1: $\left(1-Ak\right)^2 \neq 2 A^2 n \left(2+n\right)$.} In this
case the functions $P, Q, L$ are given by
\begin{equation} \label{EAnsRot2}
\begin{split}
P &= 1 + \frac{\gamma}{r^5}\,, \\
Q &= 1 + \frac{\delta}{r^5}\,, \\
L &= \frac{\epsilon}{1 + \frac{r^5}{\delta}},,
\end{split}
\end{equation}
with the constants
\begin{equation} \label{EDualConstants}
\begin{split}
\alpha &= - \frac{\left(1-Ak\right)^2 - A^2n}{\left(1-Ak\right) \left[
\left(1-Ak\right)^2 - A^2n \left(2+n\right) \right]^{1/2}}\,, \\
\beta &= \frac{An \left[ \left(1-Ak\right)^2 + A^2
\right]^{1/2}}{\left(1-Ak\right) \left[ \left(1-Ak\right)^2 - A^2n
\left(2+n\right) \right]^{1/2}}\,, \\
\gamma &= \frac{h \left(1-Ak\right)^2}{\left[ \left(1-Ak\right) + A^2
\right]^{1/6}}\,, \\
\delta &= - \frac{A^2 n^2 h \left[ \left(1-Ak\right)^2 + A^2
\right]^{5/6}}{\left(1-Ak\right)^2 - A^2 n \left(2 + n\right)}\,, \\
\epsilon &= - \frac{(1-Ak)^2 - A^2n}{An \left[ \left(1-Ak\right)^2 + A^2
\right]^{1/2}}\,.
\end{split}
\end{equation}
In fact, if one uses equations \eqref{EAnsRot1} and \eqref{EAnsRot2} as an
Ansatz for a solution to Einstein's equations one finds that they are solved
provided the constants satisfy the relations
\begin{equation} \label{RotEoMs}
\begin{split}
\alpha &= m \frac{\epsilon}{\sqrt{\epsilon^2 - 1}}\,, \\
\beta &= m \frac{1}{\sqrt{\epsilon^2 - 1}}\,, \\
\gamma &= \delta \left(1 - \epsilon^2\right)\,,
\end{split}
\end{equation}
where $m = \pm 1$. Thus we find two branches of a two-parameter family of
solutions where the free parameters are $\delta$ and $\epsilon$ and the two
branches correspond to positive and negative charges, $m = \pm 1$. It is trivial
to check that the constants given by \eqref{EDualConstants} indeed satisfy
\eqref{RotEoMs}. We note that although the dual solution has three independent
parameters $\left(A, n, k\right)$, the family of solutions only depends on two
parameters $\delta, \epsilon$. This means that some combinations of $\left(A, n,
k\right)$ give the same physical solutions. This is similar to the extreme
M2-brane where when $n = 0$ the duality always gives the same extreme M2-brane
solutions.

These solutions are delocalised along the three worldvolume directions $t, y_1, y_2$ as well as along $x_3$ and is asymptotically flat along the other seven transverse directions $z_4, \ldots, z_{10}$. The solutions carry momentum along the $x_3$ direction and upon compactifying along $x_3$ we obtain a ten-dimensional type IIA solution where this Kaluza-Klein momentum gives rise to a Ramond-Ramond 1-form, $\mathcal{A}_t$. The type IIA solution in the string frame is given by
\begin{equation}
\begin{split}
ds_{10}^2 &= - \frac{1}{\sqrt{H}} dt^2 + \sqrt{H} d\mathbf{z}^2 + \sqrt{H}
P^{-1} d\mathbf{y}^2\,, \\
B_{y_1y_2} &= \frac{\beta}{P}\,, \\
C_{ty_1y_2} &= -\frac{\alpha}{P}\,, \\
\mathcal{A}_t &= \epsilon H^{-1} \frac{\delta}{r^5}\,, \\
\phi &= \frac{3}{4} \ln H - \frac{1}{2} \ln P\,,
\end{split}
\end{equation}
where $P$ and $H$ are given by
\begin{equation}
\begin{split}
P &= 1 + \frac{\gamma}{r^5}\,, \\
H &= 1 - \frac{\epsilon^2 \delta}{r^5}\,.
\end{split}
\end{equation}

We can calculate the tension and charge densities of the solutions by Komar integrals
\begin{equation}
\begin{split}
T &= - 5 f_M \omega_{(6)} \delta \left( \frac{1}{3} + \frac{2}{3} \epsilon^2
\right)\,, \\
Q &= 5 f_Q \omega_{(6)} n \delta \epsilon \sqrt{\epsilon^2 - 1}\,,
\end{split}
\end{equation}
where the density is also over the $x_3$ coordinate.

Finally, let us mention that the singularity at $r = 0$ is not regular. The Kretschmann scalar giverges
\begin{equation}
 R_{abcd} R^{abcd} \propto r^{-2/3}\,,
\end{equation}
which is exactly the same kind of singularity as for the smeared M2-brane. Thus, the solution is singular at $r = 0$ but no more singular than our seed solution. However, in the case of the smeared M2-brane, the singularity can be resolved by noticing that the solution is the zero-mode of an array of extreme M2-branes. Including the higher mass modes, we obtain a regular solution. Thus, the smeared M2-brane should not be taken seriously near $r = 0$. We wonder whether there is a similar way to lift the singularity in the rotating case.

\paragraph{Case 2: $\left(1-Ak\right)^2 = A^2n \left(2+n\right)$} Now the metric can still be put into the form \eqref{EAnsRot1} but the functions become
\begin{equation} \label{EAnsRot3}
\begin{split}
P &= 1 + \frac{\gamma}{r^5}\,, \\
Q &= \frac{\delta}{r^5}\,, \\
L &= \epsilon + \phi r^5\,,
\end{split}
\end{equation}
with the constants given by
\begin{equation}
\begin{split}
\gamma &= n \left(2+n\right) A^{5/3} h \left(1+n\right)^{-1/3}\,, \\
\delta &= \phi = - 1\,, \\
\epsilon &= - n A^{5/3} h \left(1+n\right)^{2/3}\,, \\
\alpha &= - \frac{\left(1+n\right)^{1/6}}{A^{5/6} \sqrt{nh \left(2+n\right)}}\,,
\\
\beta &= - \frac{A^{5/6} \sqrt{nh}}{\left(1+n\right)^{1/6} \sqrt{2+n}}\,.
\end{split}
\end{equation}
One can once again take equations \eqref{EAnsRot1} and \eqref{EAnsRot3} as an
Ansatz to solve the Einstein equations. The equations of motion then restrict
the constants to be
\begin{equation}
\begin{split}
\phi &= \frac{m}{\delta}\,, \\
\gamma &= - \frac{\delta}{\alpha^2}\,, \\
\beta &= \frac{m + \alpha^2 \epsilon}{\alpha}\,,
\end{split}
\end{equation}
where $m = \pm 1$. We note that $\frac{\delta}{\gamma}$ is strictly negative. If
$\gamma < 0$ we can redefine $r \rightarrow - r$ and make it positive again.
Thus, we can without loss of generality take $\gamma > 0$ and $\delta < 0$. In
particular, this means that $g_{tt} > 0$ and thus the coordinate $t$ is not
timelike. By changing coordinates, however, we find that this solution
corresponds to a smeared M2-brane. We have to take
\begin{equation}
\begin{split}
X_3 &= \alpha \epsilon x_3 - \alpha t\,, \\
T &= -m \alpha t + m \beta x_3\,,
\end{split}
\end{equation}
to get
\begin{equation}
 \begin{split}
  ds^2 &= P^{-2/3} \left(-dT^2 + dY_1^2 + dY_2^2\right) + P^{1/3} \left(dX_3^2 + dZ_4^2 + \ldots + dZ_{10}^2\right)\,, \\
  C_{TY_1Y_2} &= m P^{-1}\,, \\
  C_{Y_1Y_2X_3} &= 0\,.
 \end{split}
\end{equation}

\section{Conclusions} \label{SConclusions}
 One of our aims was to see whether the signature of the spacetime metric may change under timelike dualities as was conjectured in \cite{Hull:1998ym,Hull:1998vg,Hull:1998br}. We found that instead of a signature change, generalised geometry forces us to include the trivector. This is not a dynamical field as it can usually be gauged away except when there are topological obstructions. We can view the difficulty of removing the trivector in the timelike case as a topological obstruction as well. We have also proven that the spacetime metric has to be of signature $\left(-,+,\ldots,+\right)$ if the generalised metric parameterises the coset
\begin{equation}
 \frac{SL(2)\times SL(3) \times GL(8)}{SO(1,1) \times SO(2,1) \times SO(8)}\,,
\end{equation}
which is the modular group of dualities acting along two spacelike and one timelike direction.\footnote{Actually, the factor $\frac{GL(8)}{SO(8)}$ is parameterised by the metric in the eight-dimensional transverse space.}

One may argue that instead of including the trivector in the dual solution where it cannot be gauged away, such dualities should not be allowed. However, these dualities do arise when we act along the worldvolume of M2-branes, which are fundamental objects of 11-dimensional supergravity, and thus seem ``natural''. Furthermore, the trivector is needed if the generalised vielbein cannot be lower triangular which arises generically when the generalised metric is Lorentzian. This is analogous to the geometric example given in section \ref{STIntSym}. The aim of generalised geometry is to treat the whole 11-dimensional supergravity geometrically, not just the spacetime metric. Thus, if we take the generalised geometry program seriously, we should proceed in the same fashion as in geometry and consider the solutions including a non-zero trivector seriously. Yet, another reason is that if we dimensionally reduce to a type IIA solution and then Buscher dualise along a single direction to obtain a type IIB solution, we find that the K\"ahler parameter of the $SL(2)$ duality group \cite{Malek:2012pw} gets mapped to a geometric $SL(2)$ in the type IIB solution, corresponding to a coordinate change \cite{Hull:1994ys}. Thus, the duality in type IIA, viewed from the perspective of type IIB \emph{is} geometric! Finally, the trivector is known to play a role in non-geometric backgrounds \cite{Grana:2008yw,Andriot:2011uh,Andriot:2012wx,Andriot:2012an,Aldazabal:2011nj,Dibitetto:2012rk} and thus we should not be deterred by the fact it arises here as well. Rather, this seems to be telling us that some geometric solutions will be linked to non-geometric ones through timelike dualities.

In section \ref{SExamples} we studied some examples of timelike $SL(2) \times SL(3)$ dualities acting on M2-brane solutions. We found that the resulting dual solutions belong to one of three types. The first type, obtained for a certain range of the duality parameter $\Omega^{t12} = A$, are dual solutions which can be expressed in the $\left(g_{11}, C_3\right)$ frame. We found that in this range the duality acts like a Harrison transformation, charging the uncharged black M2-brane and changing the charge of the extreme M2-brane. It is noteworthy that as in \cite{Malek:2012pw} we found that these new extreme M2-branes would have quantised tensions and charges if we use the quantum U-duality group $E_3(Z)$. The second type of dual solutions, obtained by transformations outside this range, include a trivector that cannot be gauged away. They need to be described in the $\left(\hat{g}_{11}, \hat{C}_3, \hat{\Omega}_3\right)$ frames where the individual bosonic fields are not uniquely defined but rather form a family of solutions lying in the orbit of the local symmetry group $SO(1,1)$. If we had extrapolated the first type of solutions to arbitrary large values of the duality parameter $A$ we would have obtained pathological solutions, for example solutions with negative tension. These are the solutions we naively found in \cite{Malek:2012pw} but we now saw that they are not obtained by duality because we ought to include a trivector. The third type of solutions were obtained by dualities at the ends of the range for which the trivector can be removed in the dual solutions. These correspond to 11-dimensional analogues of subtracted geometries \cite{Cvetic:2011dn} obtained by an ``infinite'' Harrison boost in the conventional picture. We also considered the action of $SL(5)$ on smeared M2-branes and found new solutions which contain momentum in the direction the brane is delocalised along. Because they are obtained by dualising along three spacelike directions the trivector can always be removed. The solutions contain a curvature singularity at the center of the polar coordinates, at $r=0$, of the same nature as the original smeared M2-brane solution, suggesting that there may be a stringy resolution of the singularity although this remains an open question.


Clearly, the trivector plays a fundamental role in the generalised geometry formulation of 11-dimensional supergravity. It remains an open question of how solutions including a trivector should be treated and how a M2-brane couples to these backgrounds. In particular, one may wonder what becomes of physical quantities such as mass and the 3-form charge when there is a non-zero trivector although progress has recently been made in understanding the geometry of the trivector in the context of string theory \cite{Andriot:2012an}. We wish to address these questions in a future publication.

The duality groups $SO(5,5), E_6, E_7$ and $E_8$ can be used to act on intersecting M2-branes, M5-branes and their intersections, and the Kaluza-Klein Monopole. We expect to find charging transformation when acting along their worldvolumes, including a transformation that gives a ``subtracted geometry'', and to be able to create momenta along delocalised transverse directions.

\acknowledgments
I would like to thank my supervisor, Malcolm Perry, for many helpful discussions, Garry Gibbons for insightful comments on subtracted geometries, Olaf Hohm for comments on the appearance of IIA* and IIB* string theories in Double Field Theory and Jeremy Sakstein for his feedback on the manuscript. I further wish to thank the STFC for supporting me through a Postgraduate Studentship grant and Peterhouse, Cambridge for their support through the Peterhouse Research Studentship.

\bibliographystyle{JHEP}

\begin{thebibliography}{1}
\bibitem{Cremmer:1977tt}
E.~Cremmer, J.~Scherk, and S.~Ferrara, {\it {SU(4) Invariant Supergravity
  Theory}},  {\em Phys.\ Lett.} {\bf B 74} (1978) 61.

\bibitem{Cremmer:1978ds}
E.~Cremmer and B.~Julia, {\it {The N=8 Supergravity Theory. 1. The
  Lagrangian}},  {\em Phys.\ Lett.} {\bf B 80} (1978) 48.

\bibitem{Cremmer:1979up}
E.~Cremmer and B.~Julia, {\it {The SO(8) Supergravity}},  {\em Nucl.\ Phys.}
  {\bf B 159} (1979) 141.

\bibitem{Gualtieri:2003dx}
M.~Gualtieri, {\it {Generalized complex geometry}},
  \href{http://xxx.lanl.gov/abs/math/0401221}{{\tt math/0401221}}.

\bibitem{Hitchin:2004ut}
N.~Hitchin, {\it {Generalized Calabi-Yau manifolds}},  {\em Quart.\ J.\ Math.\
  Oxford Ser.} {\bf 54} (2003) 281--308,
  [\href{http://xxx.lanl.gov/abs/math/0209099}{{\tt math/0209099}}].

\bibitem{Hitchin:2005in}
N.~Hitchin, {\it {Brackets, forms and invariant functionals}},
  \href{http://xxx.lanl.gov/abs/math/0508618}{{\tt math/0508618}}.

\bibitem{Hitchin:2005cv}
N.~Hitchin, {\it {Instantons, Poisson structures and generalized Kahler
  geometry}},  {\em Commun.\ Math.\ Phys.} {\bf 265} (2006) 131--164,
  [\href{http://xxx.lanl.gov/abs/math/0503432}{{\tt math/0503432}}].

\bibitem{Hull:2007zu}
C.~M. Hull, {\it {Generalised Geometry for M-Theory}},  {\em JHEP} {\bf 0707}
  (2007) 079, [\href{http://xxx.lanl.gov/abs/hep-th/0701203}{{\tt
  hep-th/0701203}}].

\bibitem{Berman:2010is}
D.~S. Berman and M.~J. Perry, {\it {Generalized Geometry and M theory}},  {\em
  JHEP} {\bf 1106} (2011) 074, [\href{http://xxx.lanl.gov/abs/1008.1763}{{\tt
  arXiv:1008.1763}}].

\bibitem{Berman:2011pe}
D.~S. Berman, H.~Godazgar, and M.~J. Perry, {\it {SO(5,5) duality in M-theory
  and generalized geometry}},  {\em Phys.\ Lett.} {\bf B 700} (2011) 65--67,
  [\href{http://xxx.lanl.gov/abs/1103.5733}{{\tt arXiv:1103.5733}}].

\bibitem{Berman:2011jh}
D.~S. Berman, H.~Godazgar, M.~J. Perry, and P.~C. West, {\it {Duality Invariant
  Actions and Generalised Geometry}},  {\em JHEP} {\bf 1202} (2012) 108,
  [\href{http://xxx.lanl.gov/abs/1111.0459}{{\tt arXiv:1111.0459}}].

\bibitem{Berman:2011cg}
D.~S. Berman, H.~Godazgar, M.~Godazgar, and M.~J. Perry, {\it {The Local
  symmetries of M-theory and their formulation in generalised geometry}},  {\em
  JHEP} {\bf 1201} (2012) 012, [\href{http://xxx.lanl.gov/abs/1110.3930}{{\tt
  arXiv:1110.3930}}].

\bibitem{Berman:2012vc}
D.~S. Berman, M.~Cederwall, A.~Kleinschmidt, and D.~C. Thompson, {\it {The
  gauge structure of generalised diffeomorphisms}},
  \href{http://xxx.lanl.gov/abs/1208.5884}{{\tt arXiv:1208.5884}}.

\bibitem{Berman:2012uy}
D.~S. Berman, E.~T. Musaev, and D.~C. Thompson, {\it {Duality Invariant
  M-theory: Gauged supergravities and Scherk-Schwarz reductions}},
  \href{http://xxx.lanl.gov/abs/1208.0020}{{\tt arXiv:1208.0020}}.

\bibitem{Aldazabal:2010ef}
G.~Aldazabal, E.~Andres, P.~G. Camara, and M.~Gra\~{n}a, {\it {U-dual fluxes
  and Generalized Geometry}},  {\em JHEP} {\bf 1011} (2010) 083,
  [\href{http://xxx.lanl.gov/abs/1007.5509}{{\tt arXiv:1007.5509}}].

\bibitem{Pacheco:2008ps}
P.~P. Pacheco and D.~Waldram, {\it {M-theory, exceptional generalised geometry
  and superpotentials}},  {\em JHEP} {\bf 0809} (2008) 123,
  [\href{http://xxx.lanl.gov/abs/0804.1362}{{\tt arXiv:0804.1362}}].

\bibitem{Coimbra:2011ky}
A.~Coimbra, C.~Strickland-Constable, and D.~Waldram, {\it {$E_{d(d)} \times
  R^+$ Generalised Geometry, Connections and M Theory}},
  \href{http://xxx.lanl.gov/abs/1112.3989}{{\tt arXiv:1112.3989}}.

\bibitem{Hull:2006tp}
C.~M. Hull and R.~A. Reid-Edwards, {\it {Flux compactifications of M-theory on
  twisted Tori}},  {\em JHEP} {\bf 0610} (2006) 086,
  [\href{http://xxx.lanl.gov/abs/hep-th/0603094}{{\tt hep-th/0603094}}].

\bibitem{Hull:2009mi}
C.~M. Hull and B.~Zwiebach, {\it {Double Field Theory}},  {\em JHEP} {\bf 0909}
  (2009) 099, [\href{http://xxx.lanl.gov/abs/0904.4664}{{\tt
  arXiv:0904.4664}}].

\bibitem{Hull:2009zb}
C.~M. Hull and B.~Zwiebach, {\it {The Gauge algebra of double field theory and
  Courant brackets}},  {\em JHEP} {\bf 0909} (2009) 090,
  [\href{http://xxx.lanl.gov/abs/0908.1792}{{\tt arXiv:0908.1792}}].

\bibitem{Hohm:2010pp}
O.~Hohm, C.~M. Hull, and B.~Zwiebach, {\it {Generalized metric formulation of
  double field theory}},  {\em JHEP} {\bf 1008} (2010) 008,
  [\href{http://xxx.lanl.gov/abs/1006.4823}{{\tt arXiv:1006.4823}}].

\bibitem{Hohm:2010jy}
O.~Hohm, C.~M. Hull, and B.~Zwiebach, {\it {Background independent action for
  double field theory}},  {\em JHEP} {\bf 1007} (2010) 016,
  [\href{http://xxx.lanl.gov/abs/1003.5027}{{\tt arXiv:1003.5027}}].

\bibitem{Hohm:2012gk}
O.~Hohm and B.~Zwiebach, {\it {Large Gauge Transformations in Double Field
  Theory}},  \href{http://xxx.lanl.gov/abs/1207.4198}{{\tt arXiv:1207.4198}}.

\bibitem{deWit:1986mz}
B.~de~Wit and H.~Nicolai, {\it {d = 11 supergravity with local SU(8)
  invariance}},  {\em Nucl.\ Phys.} {\bf B 274} (1986) 363.

\bibitem{West:2001as}
P.~C. West, {\it {E(11) and M theory}},  {\em Class.\ Quant.\ Grav.} {\bf 18}
  (2001) 4443--4460, [\href{http://xxx.lanl.gov/abs/hep-th/0104081}{{\tt
  hep-th/0104081}}].

\bibitem{Riccioni:2007ni}
F.~Riccioni and P.~C. West, {\it {E(11)-extended spacetime and gauged
  supergravities}},  {\em JHEP} {\bf 0802} (2008) 039,
  [\href{http://xxx.lanl.gov/abs/0712.1795}{{\tt arXiv:0712.1795}}].

\bibitem{Kleinschmidt:2003jf}
A.~Kleinschmidt and P.~C. West, {\it {Representations of G+++ and the role of
  space-time}},  {\em JHEP} {\bf 0402} (2004) 033,
  [\href{http://xxx.lanl.gov/abs/hep-th/0312247}{{\tt hep-th/0312247}}].

\bibitem{West:2003fc}
P.~C. West, {\it {E(11), SL(32) and central charges}},  {\em Phys.Lett.} {\bf
  B575} (2003) 333--342, [\href{http://xxx.lanl.gov/abs/hep-th/0307098}{{\tt
  hep-th/0307098}}].

\bibitem{West:2004kb}
P.~C. West, {\it {E(11) origin of brane charges and U-duality multiplets}},
  {\em JHEP} {\bf 0408} (2004) 052,
  [\href{http://xxx.lanl.gov/abs/hep-th/0406150}{{\tt hep-th/0406150}}].

\bibitem{West:2004iz}
P.~C. West, {\it {Brane dynamics, central charges and E(11)}},  {\em JHEP} {\bf
  0503} (2005) 077, [\href{http://xxx.lanl.gov/abs/hep-th/0412336}{{\tt
  hep-th/0412336}}].

\bibitem{Riccioni:2006az}
F.~Riccioni and P.~C. West, {\it {Dual fields and E(11)}},  {\em Phys.Lett.}
  {\bf B645} (2007) 286--292,
  [\href{http://xxx.lanl.gov/abs/hep-th/0612001}{{\tt hep-th/0612001}}].

\bibitem{Cook:2008bi}
P.~P. Cook and P.~C. West, {\it {Charge multiplets and masses for E(11)}},
  {\em JHEP} {\bf 0811} (2008) 091,
  [\href{http://xxx.lanl.gov/abs/0805.4451}{{\tt arXiv:0805.4451}}].

\bibitem{West:2010ev}
P.~C. West, {\it {$E_{11}$, generalised space-time and IIA string theory}},
  {\em Phys.Lett.} {\bf B696} (2011) 403--409,
  [\href{http://xxx.lanl.gov/abs/1009.2624}{{\tt arXiv:1009.2624}}].

\bibitem{West:2010rv}
P.~C. West, {\it {Generalised space-time and duality}},  {\em Phys.Lett.} {\bf
  B693} (2010) 373--379, [\href{http://xxx.lanl.gov/abs/1006.0893}{{\tt
  arXiv:1006.0893}}].

\bibitem{West:2011mm}
P.~C. West, {\it {Generalised geometry, eleven dimensions and E11}},  {\em
  JHEP} {\bf 1202} (2012) 018, [\href{http://xxx.lanl.gov/abs/1111.1642}{{\tt
  arXiv:1111.1642}}].

\bibitem{West:2012qz}
P.~West, {\it {E11, generalised space-time and equations of motion in four
  dimensions}},  \href{http://xxx.lanl.gov/abs/1206.7045}{{\tt
  arXiv:1206.7045}}.

\bibitem{Malek:2012pw}
E.~Malek, {\it {U-duality in three and four dimensions}},
  \href{http://xxx.lanl.gov/abs/1205.6403}{{\tt arXiv:1205.6403}}.

\bibitem{Hull:1998vg}
C.~M. Hull, {\it {Timelike T duality, de Sitter space, large N gauge theories
  and topological field theory}},  {\em JHEP} {\bf 9807} (1998) 021,
  [\href{http://xxx.lanl.gov/abs/hep-th/9806146}{{\tt hep-th/9806146}}].

\bibitem{Hull:1998ym}
C.~M. Hull, {\it {Duality and the signature of space-time}},  {\em JHEP} {\bf
  9811} (1998) 017, [\href{http://xxx.lanl.gov/abs/hep-th/9807127}{{\tt
  hep-th/9807127}}].

\bibitem{Hull:1998br}
C.~M. Hull and B.~Julia, {\it {Duality and moduli spaces for timelike
  reductions}},  {\em Nucl.\ Phys.} {\bf B 534} (1998) 250--260,
  [\href{http://xxx.lanl.gov/abs/hep-th/9803239}{{\tt hep-th/9803239}}].

\bibitem{Hohm:2011zr}
O.~Hohm, S.~K. Kwak, and B.~Zwiebach, {\it {Unification of Type II Strings and
  T-duality}},  {\em Phys.Rev.Lett.} {\bf 107} (2011) 171603,
  [\href{http://xxx.lanl.gov/abs/1106.5452}{{\tt arXiv:1106.5452}}].

\bibitem{Hohm:2011dv}
O.~Hohm, S.~K. Kwak, and B.~Zwiebach, {\it {Double Field Theory of Type II
  Strings}},  {\em JHEP} {\bf 1109} (2011) 013,
  [\href{http://xxx.lanl.gov/abs/1107.0008}{{\tt arXiv:1107.0008}}].

\bibitem{Duff:1990hn}
M.~J. Duff and J.~X. Lu, {\it {Duality Rotations in Membrane Theory}},  {\em
  Nucl.Phys.} {\bf B347} (1990) 394--419.

\bibitem{Berman:2011kg}
D.~S. Berman, E.~T. Musaev, and M.~J. Perry, {\it {Boundary Terms in
  Generalized Geometry and doubled field theory}},  {\em Phys.Lett.} {\bf B706}
  (2011) 228--231, [\href{http://xxx.lanl.gov/abs/1110.3097}{{\tt
  arXiv:1110.3097}}].

\bibitem{Andriot:2011uh}
D.~Andriot, M.~Larfors, D.~Lust, and P.~Patalong, {\it {A ten-dimensional
  action for non-geometric fluxes}},  {\em JHEP} {\bf 1109} (2011) 134,
  [\href{http://xxx.lanl.gov/abs/1106.4015}{{\tt arXiv:1106.4015}}].

\bibitem{Buscher:1987sk}
T.~H. Buscher, {\it {A Symmetry of the String Background Field Equations}},
  {\em Phys.\ Lett.} {\bf B 194} (1987) 59.

\bibitem{Buscher:1987qj}
T.~H. Buscher, {\it {Path Integral Derivation of Quantum Duality in Nonlinear
  Sigma Models}},  {\em Phys.\ Lett.} {\bf B 201} (1988) 466.

\bibitem{Grana:2008yw}
M.~Gra\~{n}a, R.~Minasian, M.~Petrini, and D.~Waldram, {\it {T-duality,
  Generalized Geometry and Non-Geometric Backgrounds}},  {\em JHEP} {\bf 0904}
  (2009) 075, [\href{http://xxx.lanl.gov/abs/0807.4527}{{\tt
  arXiv:0807.4527}}].

\bibitem{Andriot:2012wx}
D.~Andriot, O.~Hohm, M.~Larfors, D.~Lust, and P.~Patalong, {\it {A geometric
  action for non-geometric fluxes}},
  \href{http://xxx.lanl.gov/abs/1202.3060}{{\tt arXiv:1202.3060}}.

\bibitem{Andriot:2012an}
D.~Andriot, O.~Hohm, M.~Larfors, D.~Lust, and P.~Patalong, {\it {Non-Geometric
  Fluxes in Supergravity and Double Field Theory}},
  \href{http://xxx.lanl.gov/abs/1204.1979}{{\tt arXiv:1204.1979}}.

\bibitem{Aldazabal:2011nj}
G.~Aldazabal, W.~Baron, D.~Marques, and C.~Nunez, {\it {The effective action of
  Double Field Theory}},  {\em JHEP} {\bf 1111} (2011) 052,
  [\href{http://xxx.lanl.gov/abs/1109.0290}{{\tt arXiv:1109.0290}}].

\bibitem{Dibitetto:2012rk}
G.~Dibitetto, J.~J. Fernandez-Melgarejo, D.~Marques, and D.~Roest, {\it
  {Duality orbits of non-geometric fluxes}},
  \href{http://xxx.lanl.gov/abs/1203.6562}{{\tt arXiv:1203.6562}}.

\bibitem{Witten:1995ex}
E.~Witten, {\it {String theory dynamics in various dimensions}},  {\em Nucl.\
  Phys.} {\bf B 443} (1995) 85--126,
  [\href{http://xxx.lanl.gov/abs/hep-th/9503124}{{\tt hep-th/9503124}}].

\bibitem{Gueven:1992hh}
R.~Gueven, {\it {Black p-brane solutions of D = 11 supergravity theory}},  {\em
  Phys.\ Lett.} {\bf B 276} (1992) 49--55.

\bibitem{Harrison:1968}
B.~K. Harrison, {\it {New Solutions of the Einstein-Maxwell Equations from
  Old}},  {\em J.\ Math.\ Phys.} {\bf 9} (1968) 1744--1752.

\bibitem{Breitenlohner:1987dg}
P.~Breitenlohner, D.~Maison, and G.~W. Gibbons, {\it {Four-Dimensional Black
  Holes from Kaluza-Klein Theories}},  {\em Commun.Math.Phys.} {\bf 120} (1988)
  295.

\bibitem{Cvetic:2012tr}
M.~Cvetic and G.~Gibbons, {\it {Conformal Symmetry of a Black Hole as a Scaling
  Limit: A Black Hole in an Asymptotically Conical Box}},
  \href{http://xxx.lanl.gov/abs/1201.0601}{{\tt arXiv:1201.0601}}.

\bibitem{Virmani:2012kw}
A.~Virmani, {\it {Subtracted Geometry From Harrison Transformations}},
  \href{http://xxx.lanl.gov/abs/1203.5088}{{\tt arXiv:1203.5088}}.

\bibitem{Castro:2010fd}
A.~Castro, A.~Maloney, and A.~Strominger, {\it {Hidden Conformal Symmetry of
  the Kerr Black Hole}},  {\em Phys.Rev.} {\bf D82} (2010) 024008,
  [\href{http://xxx.lanl.gov/abs/1004.0996}{{\tt arXiv:1004.0996}}].

\bibitem{Cvetic:2011hp}
M.~Cvetic and F.~Larsen, {\it {Conformal Symmetry for General Black Holes}},
  {\em JHEP} {\bf 1202} (2012) 122,
  [\href{http://xxx.lanl.gov/abs/1106.3341}{{\tt arXiv:1106.3341}}].

\bibitem{Cvetic:2011dn}
M.~Cvetic and F.~Larsen, {\it {Conformal Symmetry for Black Holes in Four
  Dimensions}},  \href{http://xxx.lanl.gov/abs/1112.4846}{{\tt
  arXiv:1112.4846}}.

\bibitem{Compere:2010uk}
G.~Compere, W.~Song, and A.~Virmani, {\it {Microscopics of Extremal Kerr from
  Spinning M5 Branes}},  {\em JHEP} {\bf 1110} (2011) 087,
  [\href{http://xxx.lanl.gov/abs/1010.0685}{{\tt arXiv:1010.0685}}].

\bibitem{Bertini:2011ga}
S.~Bertini, S.~L. Cacciatori, and D.~Klemm, {\it {Conformal structure of the
  Schwarzschild black hole}},  {\em Phys.Rev.} {\bf D85} (2012) 064018,
  [\href{http://xxx.lanl.gov/abs/1106.0999}{{\tt arXiv:1106.0999}}].

\bibitem{Compere:2012jk}
G.~Compere, {\it {The Kerr/CFT correspondence and its extensions: a
  comprehensive review}},  \href{http://xxx.lanl.gov/abs/1203.3561}{{\tt
  arXiv:1203.3561}}.

\bibitem{Duff:1990xz}
M.~J. Duff and K.~S. Stelle, {\it {Multimembrane solutions of D = 11
  supergravity}},  {\em Phys.\ Lett.} {\bf B 253} (1991) 113--118.

\bibitem{Hull:1994ys}
C.~M. Hull and P.~K. Townsend, {\it {Unity of superstring dualities}},  {\em
  Nucl.\ Phys.} {\bf B 438} (1995) 109--137,
  [\href{http://xxx.lanl.gov/abs/hep-th/9410167}{{\tt hep-th/9410167}}].
\end{thebibliography}

\end{document}